\documentclass[a4paper,12pt]{article}
\usepackage{amsmath,amsfonts}
\def\ba {\begin {array}}
\def\ea{\end {array}}
\def\be {\begin {equation}}
\def\ee {\end {equation}}
\def\bea {\begin {eqnarray}}
\def\eea {\end {eqnarray}}

\begin{document}
\title
{Extended Kepler-Coulomb quantum superintegrable  systems in 3 dimensions}

\author{E. G.~Kalnins,\\
Department of Mathematics,\\
University
of Waikato, Hamilton, New Zealand,\\
J.~M.~Kress\\ 
 School of Mathematics, The University of New South Wales, \\
Sydney NSW 2052, Australia, \\
and\\
W.~Miller, Jr.\\
 School of Mathematics, University of Minnesota,\\
Minneapolis, Minnesota, U.S.A.}
\maketitle

MSC classes: 20C99, 20C35, 22E70
\begin{abstract} 
 The quantum Kepler-Coulomb  system in 3 dimensions is well known to be 2nd order superintegrable, with a symmetry algebra that
 closes polynomially under commutators. This polynomial closure is also typical for 2nd order superintegrable systems in 
2D and for 2nd order systems in 3D with nondegenerate (4-parameter) potentials.
However the degenerate 3-parameter potential for the 3D Kepler-Coulomb   system (also 2nd order superintegrable) 
is an exception,  as its  symmetry algebra doesn't close polynomially.  The 3D 4-parameter  potential for the extended  
Kepler-Coulomb  system is not even 2nd order superintegrable, but Verrier and Evans  (2008) showed  it was 4th order 
superintegrable, and   Tanoudis and Daskaloyannis (2011) showed that, if a 2nd 4th order symmetry is added to the generators,
 the symmetry algebra  closes polynomially.  Here, based on the Tremblay, Turbiner and Winternitz construction,  we  consider an infinite class of
 quantum extended Kepler-Coulomb 3 and 4-parameter systems indexed by a pair of rational numbers $(k_1,k_2)$ and 
reducing to the usual systems when $k_1=k_2=1$. We show these systems to be superintegrable of arbitrarily high order 
and determine the structure of their symmetry algebras. We demonstrate that the symmetry algebras close algebraically; only 
for systems admitting extra discrete symmetries is polynomial closure achieved. Underlying the structure theory is the 
existence of raising and lowering operators, not themselves symmetry operators or even defined independent of basis,  
that can be employed to construct the symmetry operators  and their structure relations.
\end{abstract}


\section{Introduction}\label{Intro}
A quantum  system is defined by a 
 Schr\"odinger operator 
$H=\Delta +V({\bf x})$
where $\Delta=\frac{1}{\sqrt{g}}\sum_{ij}\partial_{x_i}(\sqrt{g}g^{ij})\partial_{x_j}$ is the Laplace-Beltrami operator on an $n$-dimensional Riemannian manifold,  in local coordinates 
$x_j$. The system is maximal {\it superintegrable} of order $\ell$ if it admits $2n-1$ algebraically independent globally defined
differential  symmetry operators (the maximal number possible)
$S_j,\quad 1\le j\le 2n-1,\quad n\ge 2,$ with $S_1=H$
and $[H,S_j]\equiv HS_j-S_jH=0$, such that
$\ell$ is the maximum order of the generating symmetries (other than $H$)   as a differential operator.
 Systems associated  with Lie algebras ($\ell=1$) and separation
of variables ($\ell=2$) are the simplest and best studied. An integrable system has $n$ algebraically independent commuting
symmetry operators whereas a superintegrable system has $2n-1$
independent symmetry operators which cannot all commute and this nonabelian structure is critical for finding the
 spectral resolution of $H$ by algebraic methods alone. The importance of these systems is that they can be solved exactly. Progress in classifying and elucidating the structure of 
these systems has been impressive in the last two decades, see e.g. \cite{SCQS, Zhedanov1992a, BDK, Marquette20101} for some representative works related to the present paper.

The 3-parameter extended   Kepler Coulomb system is defined by the Hamiltonian operator 
\begin{equation}\label{KC3}{ H_3}=\partial_x^2+\partial_y^2+\partial_z^2+\frac{\alpha}{r}+\frac{\beta}{x^2}+\frac{\gamma}{y^2},\quad r=\sqrt{x^2+y^2+z^2}.\end{equation}
It is 2nd order superintegrable,  with generators $H=L_1$,
\[L_2=(x\partial_y-y\partial_x)^2+(y\partial_z-z\partial_y)^2+(z\partial_x-x\partial_z)^2+\frac{\beta r^2}{x^2}+\frac{\gamma r^2}{y^2},\]
\[ L_3=(x\partial_y-y\partial_x)^2+\frac{\beta (x^2+y^2)}{x^2}+\frac{\gamma (x^2+y^2)}{y^2},\
L_4=(x\partial_z-z\partial_x)^2+\frac{\beta z^2}{x^2+z^2},\]
\[L_5=-\frac12\{\partial_x,x\partial_z-z\partial_x\}-\frac12\{\partial_y,y\partial_z-z\partial_y\}+\frac{\alpha}{2r}+z(\frac{\beta}{x^2}+\frac{\gamma}{y^2}),\]
Here $\{A,B\} =AB+BA$.
In spherical coordinates $x=r\cos(\theta_2)\sin(\theta_1)$, $y=r\sin(\theta_2)\sin(\theta_1)$, $z=r\cos(\theta_1)$ the operators $L_2,L_3$ become
\[ L_2 =\partial_{\theta_1}^2+\cot (\theta_1)\partial_{\theta_1}+\frac{{ L}_3}{\sin^2(\theta_1)},\quad
 L_3= \partial_{\theta_2}^2+\frac{\beta}{\cos^2(\theta_2)}+\frac{\gamma}{\sin^2(\theta_2)}.\]
and we see that these are the symmetries responsible for the separation of the eigenvalue equation $H_3\Psi=E\Psi$ in spherical coordinates. Now we use the same idea as in the 
papers \cite{TTW1,TTW2} by expressing $H$ in spherical coordinates and replacing $\theta_1$ by $k_1\theta_1$ and $\theta_2$ by $k_2\theta_2$ where $k_1,k_2$ are arbitrary positive 
rational numbers. We obtain a family of  extended  3-parameter  Kepler Coulomb operators
\be\label{Hq3} { H}=\partial_r^2+\frac{2}{r}\partial_r+\frac{\alpha}{r}+\frac{1-k_1^2}{4r^2}+\frac{{ L}_2}{r^2},\ee
where 
\be\label{L23q}{ L}_2 =\partial_{\theta_1}^2+k_1\cot (k_1\theta_1)\partial_{\theta_1}+\frac{{ L}_3}{\sin^2(k_1\theta_1)},\quad
{ L}_3= \partial_{\theta_2}^2+\frac{\beta}{\cos^2(k_2\theta_2)}+\frac{\gamma}{\sin^2(k_2\theta_2)}.\ee
Again, ${ L}_2$, ${ L}_3$ are symmetry operators that determine multiplicative separation of the Schr\"odinger eigenvalue  equation $H\Psi=E\Psi$, 
and $k_j=p_j/q_j$ where $p_j,q_j$ are nonzero relatively prime positive integers for $j=1,2$, respectively. Note that
$[{ L}_2,{ L}_3]=0$ so ${ L}_2$ and ${ L}_3$ are in involution. Also note that the potential here is 
$${\tilde V}=\frac{\alpha}{r}+\frac{1-k_1^2}{4r^2}+ \frac{1}{r^2}(\frac{\beta}{\sin^2(k_1\theta_1)\cos^2(k_2\theta_2)}+
\frac{\gamma}{\sin^2(k_1\theta_1)\sin^2(k_2\theta_2)}),$$
differing from the classical potential $V$ by the additive term $\frac{1-k_1^2}{4r^2}$. This term corresponds to $R/8$ where $R$ is the scalar curvature of 
the manifold, just the correction term needed for the conformally invariant Laplacian, e.g., \cite{JUHL}, and needed here for superintegrability.
We will verify superintegrability of this system for all rational $k_1,k_2$ by explicit construction of two additional independent symmetry operators.

The 4-parameter extended Kepler Coulomb system is defined by the Hamiltonian
\begin{equation}\label{KC}{ H_4}=\partial_x^2+\partial_y^2+\partial_z^2+\frac{\alpha}{r}+\frac{\beta}{x^2}+\frac{\gamma}{y^2}+\frac{\delta}{z^2}.\end{equation}
The eigenvalue equation is separable in spherical coordinates so the system admits three commuting symmetry operators $L_1=H$, $L_2$, $L_3$, responsible for the 
separation of variables:
\be\label{L24}L_2=(x\partial_y-y\partial_x)^2+(y\partial_z-z\partial_y)^2+(z\partial_x-x\partial_z)^2+\frac{\beta r^2}{x^2}+\frac{\gamma r^2}{y^2}+\frac{\delta r^2}{z^2},\ee
\be\label{L34}L_3=(x\partial_y-y\partial_x)^2+\frac{\beta (x^2+y^2)}{x^2}+\frac{\gamma (x^2+y^2)}{y^2}.\ee
This system is not 2nd order superintegrable, but as shown in \cite{Evans2008a}, it is 4th order superintegrable. We will again verify this in \S \ref{specialcase}. 
 We  extend the system by passing to spherical coordinates and replacing each angular coordinate $\theta_i$ by $k_i\theta_i$ where $k_i$ is a fixed rational number.
The extended Kepler-Coulomb operator is  now $H\Psi=E\Psi$, where $[{ L}_2,{ L}_3]=[L_j,H]=0$ and 
\be\label{H4}{ H_4}=\partial_r^2+\frac{2}{r}\partial_r+\frac{\alpha}{r}+\frac{1-k_1^2}{4r^2}+\frac{{ L}_2}{r^2},\ee
\[{ L}_2 =\partial_{\theta_1}^2+k_1\cot (k_1\theta_1)\partial_{\theta_1}+\frac{{ L}_3}{\sin^2(k_1\theta_1)}+\frac{\delta}{\cos^2(k_1\theta_1)},\
 { L}_3= \partial_{\theta_2}^2+\frac{\beta}{\cos^2(k_2\theta_2)}+\frac{\gamma}{\sin^2(k_2\theta_2)},\]
Again  the 2nd order operators  ${ L}_2$, ${ L}_3$  are just those that determine multiplicative separation of the Schr\"odinger   equation. 
The scalar potential  is 
\[{\tilde V}=\frac{\alpha}{r}+\frac{1-k_1^2}{4r^2}+ \frac{1}{r^2}(\frac{\beta}{\sin^2(k_1\theta_1)\cos^2(k_2\theta_2)}+
\frac{\gamma}{\sin^2(k_1\theta_1)\sin^2(k_2\theta_2)}
+\frac{\delta}{\cos^2(k_1\theta_2)}).\]
It differs from the classical potential $V$ by the term $\frac{1-k_1^2}{4r^2}$ which
corresponds to $-R/8$ where $R$ is the scalar curvature of the manifold, 
Note that for $k_1\ne 1$ the space isn't flat. We will show that this system is superintegrable for all rational $k_1,k_2$.

We construct the missing symmetry operators by exploiting the following observation \cite{KKM10b,KKM10c}: The separated eigenfunctions of these Hamiltonians in spherical coordinates 
are products of functions of hypergeometric type. The
formal eigenspaces of the
Hamiltonian are invariant under action of 
any  symmetry operator, so the operator must induce recurrence relations for the
basis of separated  eigenfunctions.
Our strategy is to use the known recurrence relations 
for hypergeometric functions to reverse this process and determine a symmetry
operator from the  recurrence relations.  
We look for recurrence operators  that  change the eigenvalues of $L_2$, $L_3$ but 
preserve $E$; hence they map the eigenspace into itself. All of the special functions used in this paper are studied in \cite{AAR}, for example.

Before taking up our examples  \S \ref{canonicalform}  we describe how to compute with higher order  symmetry operators on an $n$-dimensional Riemannian or 
pseudo-Riemannian manifold with Schr\"odinger eigenvalue equation that separates multiplicatively 
in an orthogonal subgroup coordinate system. We show that such operators can be put in a canonical form which will be critical in verifying that 
differential operators that commute with $H$ on all formal eigenspaces must commute with $H$ in general.

In the following sections we show that the extended 3- and 4-parameter extended Kepler Coulomb systems are  superintegrable of arbitrarily high
 order, the order depending on $k_1,k_2$,  
and determine the structure of their symmetry algebras. We demonstrate that in general the symmetry algebras close algebraically; only 
for systems admitting extra discrete symmetries is polynomial closure achieved. Underlying the structure theory is the 
existence of raising and lowering operators, not themselves symmetry operators or defined independent of basis,  
that can be employed to construct the symmetry operators  and their structure relations. We demonstrate that our structure equations lead to two-variable models of 
 representations of the symmetry algebras in terms of difference operators. In general the eigenfunctions in these models are rational functions of 
the variables, not polynomials as in symmetry algebras with polynomial closure, e.g.  \cite{KMPost}.
We note that the proof of superintegrability for the extended 4 parameter system gives a proof of superintegrability of the 3 parameter system, 
simply by setting $\delta=0$. However, it doesn't give the minimum order generators for the 3 parameter system or the full structure. That is why we have to study each system separately.

In the special case $k_1=k_2=1$ for the 4-parameter potential we complete the results of \cite{DASK2011} and verify polynomial closure. For this system there are 6 linearly 
independent generators but we show that these generators must satisfy a symmetrized polynomial equation of order 12.

Many of the issues in this paper are quite technical.  We think that the exercise is very worth while for several reasons: 
1) The root 3 and 4-parameter systems have attracted a lot of attention during this last decade and families of extensions of them are of interest.
 2) To our knowledge this is the first explicit computation of the structure of the symmetry algebras for families of higher order superintegrable systems in 
3 variables. That it is practical to carry out the computation isn't obvious. The results show how some properties of 2nd order
 superintegrable systems and for superintegrable systems in 2 variables break down in higher dimensional cases. 3) The explicit 2-variable models 
suggest important connections with rational special functions for higher order superintegrable systems, rather than with polynomial special functions 
as for 2nd order systems.

In the related paper \cite{KM2012} we have determined the structure equations for the extended classical Kepler superintegrable systems, both 3 and 4 parameter.
The methods used to find the structure equations are very different from the quantum case but there is some similarity in the results. Classically we can prove generic 
rational closure rather than polynomial closure, whereas in the quantum case the closure is algebraic (due to noncommutivity of quantum operators). Of course the classical 
potentials must be modified to achieve quantization. In the short proceedings paper \cite{Miller2012} some of the results for the 4 parameter quantum system are announced. However, all of the 3 parameter work and 
the details of the 4 parameter computation are contained here.

\subsection{The canonical form for a 3D  symmetry operator}\label{canonicalform} 
In the special case of a 3-dimensional Riemannian or pseudo-Riemannian manifold  the defining equations for the Hamiltonian operator $ H$,
 expressed in  separable orthogonal  ``subgroup coordinates''  $q_1,q_2,q_3$, take the form \cite{KKM10}
\be \label{subgroupseparationq} { H}={ L}_1=\partial_1^2-\frac{f_1'}{f_1}\partial_1 +V_1+f_1(q_1){ L}_2,\ 
 { L}_2=\partial_2^2-\frac12\frac{f_2'}{f_2}\partial_2+V_2+f_2{ L}_3,\quad { L}_3=\partial_3^2+V_3,\ee
where the functions $f_j,V_j$ depend only on coordinate $q_j$. Thus $H=\Delta_3+{\tilde V}$ where $\Delta_3$ is the Laplace-Beltrami operator
on the manifold with metric $ds^2=dq_1^2+\frac{1}{f_1}dq_2^2+\frac{1}{f_1f_2}dq_3^2$. The operators $L_1,L_2,L_3$ are formally self-adjoint on the space with measure $dv=dq_1dq_2dq_3/\sqrt{f_1^2f_2}$. We write ${\tilde V}=V_1+f_1V_2+f_1f_2V_3$ for the scalar potential because the quantum analog of the classical Hamiltonian system with potential $V$ on the manifold may have a different scalar potential.
Thus the  Schr\"odinger eigenvalue equation is $H\Psi\equiv (\Delta_3+{\tilde V})\Psi=E\Psi$. 
Any finite order differential operator $\tilde L$ on the manifold  can be written uniquely in the standard form \cite{KKM10a}
\be\label{standardLform}
{\tilde L}=\sum_{j_1,j_2,j_3}\left(A^{j_1,j_2,j_3}({\bf q})\partial_{123}+B_1^{j_1,j_2,j_3}({\bf q})\partial_{23}+B_2^{j_1,j_2,j_3}({\bf q})\partial_{13}+B_3^{j_1,j_2,j_3}({\bf q})\partial_{12}\right.\ee
$$\left. +C_1^{j_1,j_2,j_3}({\bf q})\partial_{q_1}+C_2^{j_1,j_2,j_3}({\bf q})\partial_{q_2}+C_3^{j_1,j_2,j_3}({\bf q})\partial_{q_3}
+D^{j_1,j_2,j_3}({\bf q})
\right)L_1^{j_1}L_2^{j_2}L_3^{j_3}.$$
Note that if the formal operators  (\ref{standardLform})
contained partial
derivatives in any of $q_1,q_2,q_3$ of orders $\ge 2$ we could use the  identities (\ref{subgroupseparationq}), recursively, and 
 rearrange terms to achieve the
unique standard form (\ref{standardLform}).
In this view we can write
\be\label{generalLform}
{\tilde L}({\bf q},L_1,L_2,L_3)=A({\bf q},L_1,L_2,L_3)\partial_{123}+B_1({\bf q},L_1,L_2,L_3)\partial_{23}+B_2({\bf q},L_1,L_2,L_3)\partial_{13}\ee
\[ +B_3({\bf q},L_1,L_2,L_3)\partial_{12}+\sum_{\ell=1}^3C_\ell({\bf q},L_1,L_2,L_3)\partial_\ell+D({\bf q},L_1,L_2,L_3),\]
and consider $\tilde L$ as an at most third-order  order differential operator in $\bf q$ that is polynomial in the parameters $L_1,L_2,L_3$. Note that $A$ is the only term of 3rd order.
Then  the terms in $L_1,L_2$ and $L_3$ must be interpreted as (\ref{standardLform}) to be  considered as partial differential operators.

\subsection{Recurrence relations for the quantum Kepler-Coulomb system $H_3$}
For the system (\ref{Hq3}), which separates in spherical coordinates, $q_1=r,q_2=\theta_1,q_3=\theta_2$ we have the metric definitions
\[ f_1(r)=\frac{1}{r^2},\ V_1(r)= \frac{\alpha}{r}+\frac{1-k_1^2}{4r^2},\ f_2(\theta_1)=\frac{1}{\sin^2(k_1\theta_1)}, V_2(\theta_1)=0,\ 
V_3(\theta_2)=\frac{\beta}{\cos^2(k_2\theta_2)}+\frac{\gamma}{\sin^2(k_2\theta_2)}.\]
The difference between the quantum and classical potentials is the term $\frac{1-k_1^2}{4r^2}=-\frac{{\cal R}}{8}$ where ${\cal R}$ is the scalar curvature of 
the manifold. This is 
the quantization condition; it is 0 for flat space $k_1=1$. The separation equations for the equation $H\Psi=E\Psi$ are:
\[ 1)\quad (\partial_{\theta_2}^2+\frac{k_2^2(\frac14- b^2)}{\cos^2(k_2\theta_2)}+\frac{k_2^2(\frac14- c^2)}{\sin^2(k_2\theta_2)})\Phi(\theta_2)=-\mu^2\Phi(\theta_2),
 \quad L_3\Psi=-\mu^2\Psi,\]
where  $\beta =k_2^2(\frac14- b^2), \gamma= k_2^2 (\frac14-c^2)$, and  $\mu=k_2(2m+b+c+1)$;
\[ 2)\quad (\partial_{\theta_1}^2 +k_1\cot(k_1\theta_1)\partial_{\theta_1}-\frac{\mu^2}{\sin^2(k_1\theta_1)})\Theta(\theta_1)=-k_1^2\nu(\nu+1)\Theta(\theta_1),\quad
 L_2\Psi=-k_1^2\nu(\nu+1)\Psi,\]
where we write $\nu=(\rho-1)/2$;
\[ 3)\quad (\partial _r^2+\frac{2}{r}\partial_r + \frac{\frac{1-k_1^2\rho^2}{4}}{r^2}+\frac{\alpha}{r}-E)R(r)=0, \quad H\Psi=E\Psi.\]
In 3) it is convenient to introduce a new variable ${\rm R}=2\sqrt{E}r$ so that the separation equation becomes
\[3)' \quad (\partial _{{\rm R}}^2 + \frac{\frac{1-k_1^2\rho^2}{4}}{{\rm R}^2}+\frac{\alpha}{2\sqrt{E}{\rm R}}-\frac14)S({\rm R})=0,\quad
R({\rm R})=S(r)/{\rm R}.\]
The separated solutions are $\Psi^{n,\rho,m}=R^{k_1\rho}_n(r)\Phi_m^{(c,b)}(\cos(2k_2\theta_2)) \Theta^m_d(\cos(k_1\theta_1))$ with
$$\Phi_m^{(c,b)}(\cos(2k_2\theta_2))=\sin^{c+1/2}(k_2\theta_2)\cos^{b+1/2}(k_2\theta_2)P_m^{(c,b)}(\cos(2k_2\theta_2)),$$
where the $P_m^{(c,b)}(\cos(2k_2\theta_2))$ are Jacobi functions;
$$ \Theta^m_d(\cos(k_1\theta_1))=P^{\mu/k_1}_\nu(\cos(k_1\theta_1)),$$
where the $P^{\mu/k_1}_\nu(\cos(k_1\theta_1)),$  are associated Legendre  functions; and 
$$ R^{k_1\rho}_n(r)=\frac{S(r)}{2\sqrt{E}r},\quad S(r)=(2\sqrt{E})^{k_1\rho/2}e^{-\sqrt{E}r}r^{(k_1\rho+1)/2}L_n^{k_1\rho}(2\sqrt{E}r),$$
where the $L_n^{k_1\rho}(2\sqrt{E}r)$ are associated Laguerre functions, and the relation between $E$ and $n$ is the quantization condition 
\be\label{E_n} E=\frac{\alpha^2}{(2n+k_1\rho+1)^2}.\ee
Note: We are only interested in the space of generalized eigenfunctions, not the normalization of any individual eigenfunction. 
Thus the  relations to follow are valid on generalized eigenspaces and don't necessarily agree with the normalization of, say,  common polynomial eigenfunctions.

Now we look for recurrence operators acting on the separated eigenfunctions that will change the eigenvalues of $L_2$ and $L_3$ but preserve $E$.
{}From (\ref{E_n}) we see that $E$ is preserved under either of the transformations:
$$ n\to n+p_1,\  \rho\to \rho-2q_1\qquad {\rm or} \qquad n\to n-p_1,\ \rho\to \rho+2q_1.$$
Writing $S(r)=S^{k_1\rho}_n({\rm R})$, we can use recurrence relations for associated Laguerre functions \cite{AAR}  to obtain
\[ [2(k_1\rho+1)\partial_{\rm R}+(2n+k_1\rho+1)-\frac{(k_1\rho+1)^2}{\rm R}  )]    S^{k_1\rho}_n({\rm R}) = -2S^{k_1\rho+2}_{n-1}({\rm R}),\]
\[ [2(-k_1\rho+1)\partial_{\rm R}+(2n+k_1\rho+1)-\frac{(k_1\rho-1)^2}{\rm R})]
S^{k_1\rho}_n({\rm R}) =-2(n+1)(n+k_1\rho)S_{n+1}^{k_1\rho-2}({\rm R}).\]
Recalling that ${\rm R}=2r\alpha/(2n+k_1\rho+1)$, we see that these relations can be written as 
\bea\label{Dnpmtilde} {\tilde Y(1)}^n_-T^{k_1\rho}_n(r)&=&[2(k_1\rho+1)\partial_r+(2\alpha-\frac{(k_1\rho+1)^2}{r})]T_n^{k_1\rho}(r)=-\frac{4\alpha}{2n+k_1\rho+1}T^{k_1\rho+2}_{n-1}(r),\\
{\tilde  Y(1)}^n_+T^{k_1\rho}_n(r)&=&[2(-k_1\rho+1)\partial_r+(2\alpha-\frac{(k_1\rho-1)^2}{r})]T_n^{k_1\rho}(r)=-\frac{4\alpha}{2n+k_1\rho+1}(n+1)(n+k_1\rho)T^{k_1\rho-2}_{n+1}(r),\nonumber\eea
where $T^{k_1\rho}_n(r)=S^{k_1\rho}_n({\rm R})$. To get the action of the recurrences on the separated solutions $R(r)^{k_1\rho}_n(r)=S(r)/(2\sqrt{E}r)$ we have to perform
 the gauge transformation of multiplying by ${\rm R}^{-1}$ so that the recurrences transform as $r^{-1}{\tilde Y(1)}^n_+r={Y(1)}^n_+$,
 $r^{-1}{\tilde Y(1)}^n_-r={Y(1)}^n_-$.
Note that the factor $\sqrt{E}$ cancels out of the operator transformation formulas since the recurrences preserve energy. We obtain
\bea\label{Dnpmt} { Y(1)}^n_-R^{k_1\rho}_n(r)&=&[2(k_1\rho+1)\partial_r+(2\alpha+\frac{1-k_1^2\rho^2}{r})]R_n^{k_1\rho}(r)=-\frac{4\alpha}{2n+k_1\rho+1}R^{k_1\rho+2}_{n-1}(r),\\
{  Y(1)}^n_+R^{k_1\rho}_n(r)&=&[2(-k_1\rho+1)\partial_r+(2\alpha+\frac{1-k_1^2\rho^2}{r})]R_n^{k_1\rho}(r)=-\frac{4\alpha}{2n+k_1\rho+1}(n+1)(n+k_1\rho)
R^{k_1\rho-2}_{n+1}(r).\nonumber\eea
We can also find raising and lowering operators for the $\theta_1$-dependent part of the separable solutions, using differential recurrence relations for 
associated Legendre functions 
\bea\label{Jpm} X(1)_+^dP^\mu_{\frac{\rho-1}{2}}(x)&=&((1-x^2)\partial_x+\frac{-\rho-1}{2}x)P^\mu_{\frac{\rho-1}{2}}(x)=(\frac{\rho+1}{2}-\mu)P^\mu_{\frac{\rho+1}{2}}(x),\\
X(1)_-^dP^\mu_{\frac{\rho-1}{2}}(x)&=&((1-x^2)\partial_x+ \frac{\rho-1}{2}x)P^\mu_{\frac{\rho-1}{2}}(x)=(\frac{\rho-1}{2}+\mu)P^\mu_{\frac{\rho-3}{2}}(x),\nonumber\eea
where $x=\cos(k_1\theta_1)$. 
 We now  construct the two operators
\begin{equation}\label{recurrence3} J^+=\left(Y(1)_+^{n+(p_1-1)}Y(1)_+^{n+(p_1-2)}\cdots Y(1)_+^{n+1}Y(1)_+^n\right)
\left(X(1)^{\rho-2(q_1-1)}_-X(1)^{\rho-2(q_1-2)}_- \cdots X(1)^{\rho-2}_-X(1)^\rho_-\right),\end{equation}
\begin{equation}\label{recurrence4}J ^-=\left(Y(1)_-^{n-(p_1-1)}Y(1)_-^{n-(p_1-2)}\cdots
Y(1)_-^{n-1}Y(1)_-^n\right)\left(X(1)_+^{\rho+2(q_1-1)}X(1)^{\rho+2(q_1-2)}_+\cdots X(1)^{\rho+2}_+X(1)^\rho_+\right), \end{equation}
each with $p_1+q_1$ factors. 
When applied to a basis function $\Psi^{n,\rho,m}\equiv\Psi_n$ for fixed
$2n+k_1\rho$ and $m$,
these operators  raise and lower indices according to
\begin{eqnarray}\label{raise1} J^+\Psi_n&=&\left(\frac{4\alpha}{u}\right)^{p_1}(-\frac{u}{2}-\frac{k_1\rho}{2}+\frac12)_{p_1}(\frac{u}{2}-\frac{k_1\rho}{2}+\frac12)_{p_1}(-\frac{\rho}{2}-\frac{\mu}{k_1}+\frac12)_{q_1}\Psi_{n+p_1},\\
\label{lower1} J ^-\Psi_n&=&\left(\frac{4\alpha}{u}\right)^{p_1}(\frac{\rho}{2}-\frac{\mu}{k_1}+\frac12)_{q_1}\Psi_{n-p_1},\quad u=2n+k_1\rho+1,
\end{eqnarray}
where $(\alpha)_q=(\alpha)(\alpha+1)\cdots(\alpha+q-1)$ for nonnegative integer
$q$. 

Similarly we can raise and lower $m$ while fixing $\rho$ and $n$, thus preserving the energy eigenspace: The recurrences 
for  $x=\cos(k_1\theta_1)$,
\bea Y(2)_+^m\Theta^m_\rho(x)&=&\left[\sqrt{1-x^2}\frac{d}{dx}+\frac{\mu}{k_1}\frac{ x}{\sqrt{1-x^2}}\right]\Theta^m_\rho(x)=-\Theta^{m+k_1/2k_2}_\rho(x),\nonumber\\
 Y(2)_-^m\Theta^m_\rho(x)&=&\left[\sqrt{1-x^2} \frac{d}{dx}-\frac{\mu}{k_1} \frac{ x}{\sqrt{1-x^2}}\right]\Theta^m_\rho(x)=\left(\frac{\rho-1}{2}+\frac{\mu}{k_1}\right)
\left(\frac{\rho+1}{2}-\frac{\mu}{k_1}\right)  \Theta^{m-k_1/2k_2}_\rho(x),\nonumber\eea
and the recurrences for $z=\cos(2k_2\theta_2)$,
\be\label{X2} X(2)_+^m \Phi^{(c,b)}_m(z)\equiv\ee
\[\left[(1+\frac{\mu}{k_2})(1-z^2)\frac{d}{dz}-\frac{\mu}{2k_2}(1+\frac{\mu}{k_2})z-\frac12(c^2-b^2)\right]\Phi^{(c,b)}_m(z)
=\frac12(\frac{\mu}{k_2}-b-c+1)(\frac{\mu}{k_2}+b+c+1\Phi^{(c,b)}_{m+1}(z),\]
$$ X(2)_-^m \Phi^{(c,b)}_m(z)\equiv$$
\[\left[ (1-\frac{\mu}{k_2})(1-z^2)\frac{d}{dz}+\frac{\mu}{2k_2}(1-\frac{\mu}{k_2})z-\frac12(c^2-b^2)\right]\Phi^{(c,b)}_m(z)
=\frac12(\mu-b+c-1)(\mu+b-c-1)\Phi^{(c,b)}_{m-1}(z).\]
We  construct the two operators
\begin{equation}\label{recurrence5} K^+=\left(Y(2)_+^{m+(2q_1p_2-1)k_1/2k_2}Y(2)_+^{m+(2q_1p_2-2)k_1/2k_2}\cdots Y(2)_+^{m+k_1/2k_2}Y(2)_+^m\right)
\times\ee
$$\left(X(2)^{m+p_1q_2-1}_+X(2)^{m+p_1q_2-2}_+ \cdots X(2)^{m+1}_+X(2)^m_+\right),$$
\begin{equation}\label{recurrence6}K ^-=\left(Y(2)_-^{m-(2q_1p_2-1)k_1/2k_2}Y(2)_-^{m-(2q_1p_2-2)k_1/2k_2}\cdots
Y(2)_-^{m-k_1/2k_2}Y(2)_-^m\right)\times\ee
$$\left(X(2)_-^{m-p_1q_2+1}X(2)^{m-p_1q_2+2}_-\cdots X(2)^{m+1}_-X(2)^m_-\right), $$
each with $p_1q_2+2p_2q_1$ factors. 
When applied to a basis function $\Psi^{n,\rho,m}\equiv\Psi^m$ for fixed
$n$ and $\rho$,
these operators  raise and lower indices according to
\begin{eqnarray}\label{raise2} K^+\Psi^m&=&(-2)^{p_1q_2}(\frac{\mu}{2k_2}-\frac{b}{2}-\frac{c}{2}+\frac12)_{p_1q_2}(\frac{\mu}{2k_2}+\frac{b}{2}+\frac{c}{2}+\frac12)_{p_1q_2}\Psi^{m+p_1q_2},\\
\label{lower2} K ^-\Psi^m&=&
2^{p_1q_2}(\frac{1-\rho}{2}-\frac{\mu}{k_1})_{2p_2q_1}
(\frac{1+\rho}{2}-\frac{\mu}{k_1})_{2p_2q_1}(-\frac{\mu}{2k_2}+\frac{b}{2}-\frac{c}{2}+\frac12)_{p_1q_2}(-\frac{\mu}{2k_2}
-\frac{b}{2}+\frac{c}{2}+\frac12)_{p_1q_2}\Psi^{m-p_1q_2}.\nonumber
\end{eqnarray}
Note that each of the separated solutions $R^{k_1\rho}_n(r),\Theta^m_d(x),\Phi^{(c,b)}_m(z)$ is characterized as an eigenfunction of a second order ordinary 
differential operator. Hence, except for isolated values of the parameters  it is possible to find bases 
$$  (R^{k_1\rho}_n(r),{\tilde R}^{k_1\rho}_n(r)),\ (\Theta^m_\rho(x),{\tilde \Theta}^m_\rho(x)),\ (\Phi^{(c,b)}_m(z),\ ({\tilde \Phi}^{(c,b)}_m(z))$$
for each of these eigenspaces such that each element of the basis set satisfies exactly the same recurrence formulas. In particular, 
the solution space of $H\Psi =E\Psi$  is  8 dimensional.

{}From the explicit expressions (\ref{recurrence3}), (\ref{recurrence4}) for the $J^\pm$
operators it is easy to verify that 
under the transformation $\rho \rightarrow  -\rho$  we have $J^+ \rightarrow 
J^-$ and $J^- \rightarrow  J^+$. Then we see that 
 \be\label{J1J2}J_1=\frac{J^--J^+}{\rho}\quad  {\rm and}\quad  J_2=J^-+J^+\ee
  are  each unchanged under this
transformation. Note: When applied to an eigenbasis we interpret $J_1$ ordered as $(J^--J^+)\frac{1}{\rho}$.
Therefore each is a polynomial in $\rho^2$.   As a consequence of 
the eigenvalue equation $L_2\Psi=-\frac{k_1^2}{4}(\rho^2-1)\Psi$,   in the
expansions of 
$J_1,J_2$ in terms of powers of $\rho^2$ we can replace $\rho^2$
by $1-4L_2/k_1^2$ everywhere it occurs, and express $J_1,J_2$ as 
pure differential operators, independent of parameters. 
Clearly each of these operators commutes with $H$ on the
eigenspaces of $H$.

Under the reflection $\mu\rightarrow  -\mu$  we have $K^+ \rightarrow 
K^-$ and $K^- \rightarrow  K^+$. Thus
 \be \label{K1K2} K_1=\frac{K^--K^+}{\mu}\quad {\rm and}\quad  K_2=K^-+K^+\ee
  are  each unchanged under this
transformation. 
Therefore each is a polynomial in $\mu^2$,   As a consequence of 
the eigenvalue equation $L_3\Psi=-\mu^2\Psi$,   in the
expansions of 
$K_1,K_2$ in terms of even powers of $\mu$ we can replace $\mu^2$
by $-L_3$ everywhere it occurs, and express $K_1,K_2$ as 
pure differential operators, independent of parameters. 
Again each of these operators commutes with $H$ on its
eigenspaces.

We have now constructed  partial differential operators $J_1,J_2,K_1,K_2$, each of which commutes with the 
Hamiltonian $H$ on all its 8-dimensional formal eigenspaces. Thus they act like symmetry operators. 
However, to prove that they are true symmetry operators  we must show that they commute 
with $H$ when acting on {\it any} analytic functions, not just separated eigenfunctions. To
establish this fact we use the canonical form of Section \ref{canonicalform}.
To be specific, we  consider the commutator $[H,J_1]$. When acting on the 8-dimensional space of formal eigenfunctions
$\Psi _0$ of $H$ the commutator gives 0. 
We want to show that it vanishes identically. To do this we write  $[H,J_1]$
in the canonical form (\ref{generalLform}): 
\be\label{4eqns4unkowns}[H,J_1 ]\Psi _0 =\left(A\partial_{123}+B_1\partial_{23}+B_2\partial_{13}+B_3\partial_{12}
+\sum_{\ell=1}^3C_\ell\partial_\ell+D\right)\Psi _0=0.\ee
Noting that we have  8 linearly
independent choices for $\Psi _0=R^{k_1\rho}_n\Theta^m_\rho\Psi^{(c,b)}_n$,
 we can express
(\ref{4eqns4unkowns}) as a system of 8 homogeneous equations for the 8 unknowns
$A,B_1,B_2,B_3,C_1,C_2,C_3,D$. We write this system in the matrix form ${\cal V}v=0$ where
$v^{\rm tr}=(A,B_1,B_2,B_3,C_1,C_2,C_3,D)$
and $V$ is an $8\times 8$ matrix built out of  the possible products of the 3 basis eigenfunctions and their first derivatives. 
By a straightforward computation we can show that
\be\label{wronskian} \det {\cal V}=\pm W(R^{k_1\rho}_n,{\tilde R}^{k_1\rho}_n)^4W(\Theta^m_\rho,{\tilde \Theta}^m_\rho)^4W(\Phi^{(c,b)}_m,{\tilde \Phi}^{(c,b)}_m)^4\ne 0\ee
where  $W(R^{k_1\rho}_n,{\tilde R}^{k_1\rho}_n)\ne 0$ is the Wronskian of the two basis solutions.
Thus we conclude that  $A=B_j=C_j=D=0$. Consequently,  $[H,J_1 ]=0$ identically. A similar argument shows that $J_2,K_1,K_2$ are also true symmetry operators. We will work out the structure of the symmetry algebra generated by $H,L_2,L_3,J_1,K_1$ and from this it will be clear that the system is superintegrable.

To determine the structure relations it is sufficient to establish them on the generalized eigenbases. Then an argument analogous to 
 the treatment of (\ref{4eqns4unkowns}) shows that the relations hold when operating on general analytic functions. We start by using the 
definitions (\ref{J1J2}) and computing on an eigenbasis. It is straightforward to verify the relations: 
\[[J_1,L_2]=k_1^2q_1J_2+k_1^2q_1^2J_1,\ 
 [J_2,L_2]=-2q_1\{J_1,L_2\}-p_1^2J_2-k_1^2q_1(1+2q_1^2)J_1,\
[J_1,L_3]=[J_2,L_3]=0.\]
Here $\{A,B\}=AB+BA$ is the anticommutator.  We set $u=2n+k_1\rho+1$ and note that $E=\alpha^2/u^2$.  Making use of the relation
$(-c)_q=(-1)^q(c-q+1)_q$ we can obtain
$$J^-J^+\Psi_n=\left(\frac{4\alpha}{u}\right)^{2p_1}(-\frac{\rho}{2}+\frac{\mu}{k_1}+\frac12)_{q_1}
(-\frac{\rho}{2}-\frac{\mu}{k_1}+\frac12)_{q_1}(\frac{u}{2}-\frac{k_1\rho}{2}+\frac12)_{p_1}(-\frac{u}{2}-\frac{k_1\rho}{2}+\frac12)_{p_1}\Psi_n$$
$$\equiv\xi_n\Psi_n,$$
$$J^+J^-\Psi_n=\left(\frac{4\alpha}{u}\right)^{2p_1}(\frac{\rho}{2}+\frac{\mu}{k_1}+\frac12)_{q_1}
(\frac{\rho}{2}-\frac{\mu}{k_1}+\frac12)_{q_1}(-\frac{u}{2}+\frac{k_1\rho}{2}+\frac12)_{p_1}(\frac{u}{2}+\frac{k_1\rho}{2}+\frac12)_{p_1}\Psi_n$$
$$\equiv\eta_n\Psi_n.$$
Note that each of $\xi_n$, $\eta_n$ is invariant under the transformation $u\to -u$, hence each is a polynomial in $1/u^2\sim E$. Similarly, each of $\xi_n$, $\eta_n$ 
is invariant under the transformation $\mu\to -\mu$, so each is a polynomial in  $\mu^2\sim L_3$.
 Further, under
 the transformation $\rho\to -\rho$ we see that $\xi_n$, $\eta_n$ switch places. Thus $\xi_n+\eta_n$ is a polynomial in each of  $1/u^2$, $\mu^2$ and $\rho^2$, hence a 
polynomial in $H$, $L_2$ and $L_3$:
$$J^+J^-+J^-J^+ =P^+(H,L_2,L_3).$$
Here the coefficients of the polynomial $P^+$ depend on the parameters $\alpha,b,c, p_1,q_1,p_2,q_2$. Thus the basis dependent operator $J^+J^-+J^-J^+$ extends to a true globally defined 
symmetry operator $P^+$.

By a similar argument,  $(\xi_n-\eta_n)/d$ is invariant under the transformation $\rho\to -\rho$, hence  a polynomial in $\rho^2$. Thus 
$$\frac{J^+J^--J^-J^+}{\rho} =P^-(H,L_2,L_3)$$
and the basis dependent operator $(J^+J^--J^-J^+)/\rho$ extends to a true globally defined 
symmetry operator $P^-$, a polynomial in  $H,L_2,L_3$. 
Then direct computation gives the structure equations
$$[J_1,J_2]=-2q_1J_1^2-2P^-.$$ 
Interim results are $J_2^2=J_1^2(1-\frac{4}{k_1^2}L_2)+2P^++2q_1J_1J_2$, and the symmetrizations 
 $\{J_1,J_1,L_2\}=6J_1^2L_2-4p_1^2J_1^2-3k_1^2q_1\{J_1,J_2\}+2k_1^2q_1P^-$, $J_1J_2=\frac12\{J_1,J_2\}-q_1J_1^2-P^-$. Thus we have the symmetrized form
$$J_2^2=(1-\frac{14q_1^2}{3})J_1^2-\frac{2}{3k_1^2}\{J_1,J_1,L_2\}-q_1\{J_1,J_2\}-\frac{2q_1}{3}P^-+2P^+.$$
Here $\{A,B,C\}=ABC+ACB+BAC+BCA+CAB+CBA$.

Similarly, for the $K$ operators 
it is straightforward to verify: 
\[ [K_1,L_3]=4p_1^2p_2^2K_1-4p_1p_2K_2,\
 [K_2,L_3]=2p_1p_2k_2\{K_1,L_3\}+8p_1^3p_2^3k_2K_1-4k_2^2p_1^2p_2q_2K_2,\
[K_1,L_2]=[K_2,L_2]=0.\] 
We find
\[ K^+K^-\Psi^m=\] \[(-4)^{p_1q_2}(-\frac{\mu}{2k_2}+\frac{b}{2}+\frac{c}{2}+\frac12)_{p_1q_2}
(-\frac{\mu}{2k_2}+\frac{b}{2}-\frac{c}{2}+\frac12)_{p_1q_2}(-\frac{\mu}{2k_2}-\frac{b}{2}+\frac{c}{2}+\frac12)_{p_1q_2}
(-\frac{\mu}{2k_2}-\frac{b}{2}-\frac{c}{2}+\frac12)_{p_1q_2}\]
\[\times (\frac{1-\rho}{2}-\frac{\mu}{k_1})_{2p_2q_1}(\frac{1+\rho}{2}-\frac{\mu}{k_1})_{2p_2q_1}\Psi^m
\equiv\zeta_m\Psi_m,\]
\[ K^-K^+\Psi^m=\]
\[(-4)^{p_1q_2}(\frac{\mu}{2k_2}-\frac{b}{2}-\frac{c}{2}+\frac12)_{p_1q_2}(\frac{\mu}{2k_2}-\frac{b}{2}+\frac{c}{2}+\frac12)_{p_1q_2}
(\frac{\mu}{2k_2}+\frac{b}{2}-\frac{c}{2}+\frac12)_{p_1q_2}(\frac{\mu}{2k_2}+\frac{b}{2}+\frac{c}{2}+\frac12)_{p_1q_2}\]
\[\times (\frac{1+\rho}{2}+\frac{\mu}{k_1})_{2p_2q_1}(\frac{1-\rho}{2}+\frac{\mu}{k_1})_{2p_2q_1}\Psi^m
\equiv \omega_m\Psi_m.\]
Note that each of $\zeta_m$, $\omega_m$ is invariant under the transformation $\rho\to -\rho$, hence each is a polynomial in $\rho^2\sim L_2$. 
 Further, under
 the transformation $\mu\to -\mu$ we see that $\zeta_m$, $\omega_m$ switch places. Thus $\zeta_m+\omega_m$  is a polynomial in each of 
 $\mu^2$  and $\rho^2$, hence a 
polynomial in  $L_3$ and $L_2$:
$$K^+K^-+K^-K^+ =S^+(L_2,L_3).$$
Again the coefficients of the polynomial $S^+$ depend on the parameters $\alpha,b,c, p_1,q_1,p_2,q_2$. Thus the basis dependent operator $K^+K^-+K^-K^+$ extends
 to a true globally defined 
symmetry operator $S^+$. 

Similarly,  $(\zeta_m-\omega_m)/\mu$  is a polynomial in each of 
 $\mu^2$  and $\rho^2$, hence also a 
polynomial in  $L_3$ and $L_2$:
$$\frac{K^+K^--K^-K^+}{\mu} =S^-(L_2,L_3).$$ Thus the basis dependent operator $(K^+K^--K^-K^+)/\mu$ extends
 to the  true globally defined 
symmetry operator $S^-$.
Additional computations yield
\[ [K_1,K_2]=2p_1p_2K_1^2-2S^-,\
 K_1K_2=\frac12\{K_1,K_2\}+p_1p_2K_1^2-S^-,\
-K_1^2L_3=K_2^2-2S^++2p_1p_2K_1K_2,\]
\[\{K_1,K_1,L_3\}=6K_1^2L_3-24p_1^2p_2^2K_1^2+12p_1p_2\{K_1,K_2\}+8k_2^2p_1^2p_2q_2K_1^2-8p_1p_2S^-.\]
The last three identities are needed for the symmetrized result
$$ K_2^2=\frac{1}{6}\{K_1,K_1,L_3\}+4k_2^2p_1^2q_2(q_2-\frac{p_2}{3})K_1^2-3p_1p_2\{K_1,K_2\}+\frac{10}{3}p_1p_2S^-+2S^+.$$
We write 
\[ J^+\Psi^{n,\rho,m}={\cal J}^+(n,\rho,m)\Psi^{n+p_1,\rho-2q_2,m},\quad
J^-\Psi^{n,\rho,m}={\cal J}^-(n,\rho,m)\Psi^{n-p_1,\rho+2q_2,m},\]
\[ K^+\Psi^{n,\rho,m}={\cal K}^+(m)\Psi^{n,\rho,m+p_1q_2},\quad 
 K^-\Psi^{n,\rho,m}={\cal K}^-(m,\rho)\Psi^{n,\rho,m-p_1q_2},\]
where ${\cal J}^\pm,{\cal K}^\pm$ are defined by the right hand sides of equations (\ref{raise1}),(\ref{lower1}),  (\ref{raise2}),(\ref{lower2}).
Then
\[[J^+,K^+]\Psi_{n,\rho,m}=\frac{1-A(\rho,\mu,)}{1+A(\rho,\mu,)}\{J^+,K^+\}\Psi_{n,\rho,m}\equiv C(\rho,\mu)\{J^+,K^+\}\Psi_{n,\rho,m},\]
\[[J^-,K^+]\Psi_{n,\rho,m}=\frac{1-A(-\rho,\mu,)}{1+A(-\rho,\mu,)}\{J^-,K^+\}\Psi_{n,\rho,m}\equiv C(-\rho,\mu)\{J^-,K^+\}\Psi_{n,\rho,m},\]
\[[J^+,K^-]\Psi_{n,\rho,m}=\frac{1-A(\rho,-\mu,)}{1+A(\rho,-\mu,)}\{J^+,K^-\}\Psi_{n,\rho,m}\equiv C(\rho,-\mu)\{J^+,K^-\}\Psi_{n,\rho,m},\]
\[[J^-,K^-]\Psi_{n,\rho,m}=\frac{1-A(-\rho,-\mu,)}{1+A(-\rho,-\mu,)}\{J^-,K^-\}\Psi_{n,\rho,m}\equiv C(-\rho,-\mu)\{J^-,K^-\}\Psi_{n,\rho,m},\]
where
$$A(\rho,\mu)=
\frac{(-\frac{\rho}{2}-\frac{\mu}{k_1}+\frac12)_{q_1}}{(-\frac{\rho}{2}-\frac{\mu}{k_1}-2q_1p_2+\frac12)_{q_1}}.$$
From this we can compute the relations
$$[K_2,J_2]=Q^{22}_{11}\{J_1,K_1\}+Q^{22}_{12}\{J_1,K_2\}+Q^{22}_{21}\{J_2,K_1\}+Q^{22}_{22}\{J_2,K_2\},$$
where the $Q^{22}_{j\ell}$ are rational functions of $\rho^2$ and $\mu^2$. 
In particular,
$$ Q^{22}_{22}=\frac14\left(C(\rho,\mu)+C(-\rho,\mu)+C(\rho,-\mu)+C(-\rho,-\mu)\right),$$
$$Q^{22}_{11}=\frac{\rho\mu}{4}\left(C(\rho,\mu)-C(-\rho,\mu)-C(\rho,-\mu)+C(-\rho,-\mu)\right),$$
$$Q^{22}_{12}=\frac{\rho}{4}\left(-C(\rho,\mu)+C(-\rho,\mu)-C(\rho,-\mu)+C(-\rho,-\mu)\right),$$
$$ Q^{22}_{21}=\frac{\mu}{4}\left(-C(\rho,\mu)-C(-\rho,\mu)+C(\rho,-\mu)+C(-\rho,-\mu)\right).$$
Similarly we find
$$[K_1,J_1]=Q^{11}_{11}\{J_1,K_1\}+Q^{11}_{12}\{J_1,K_2\}+Q^{11}_{21}\{J_2,K_1\}+Q^{11}_{22}\{J_2,K_2\},$$
$$[K_2,J_1]=Q^{12}_{11}\{J_1,K_1\}+Q^{12}_{12}\{J_1,K_2\}+Q^{12}_{21}\{J_2,K_1\}+Q^{12}_{22}\{J_2,K_2\},$$
$$[K_1,J_2]=Q^{21}_{11}\{J_1,K_1\}+Q^{21}_{12}\{J_1,K_2\}+Q^{21}_{21}\{J_2,K_1\}+Q^{21}_{22}\{J_2,K_2\},$$
where the $Q^{ij}_{k\ell}$ are rational functions of  $\rho^2$ and $\mu^2$. These $Q$ functions are related to one another. Most particularly,
$ Q^{11}_{11}=Q^{12}_{12}=Q^{21}_{21}= Q^{22}_{22}$.
These relations make sense only on a generalized eigenbasis, but can be cast into the pure operator form 
$$[K_\ell ,J_h] Q=\{J_1,K_1\}P^{h\ell}_{11}+\{J_1,K_2\}P^{h\ell}_{12}+\{J_2,K_1\}P^{h\ell}_{21}+\{J_2,K_2\}P^{h\ell}_{22},\quad h,\ell=1,2,$$
where $Q$ and $P^{h\ell}_{jk}$ are polynomial operators in $L_2,L_3$. In particular, $Q$ is the symmetry operator defined by the product
$ B(\rho,\mu)B(\rho,-\mu)B(-\rho,\mu)B(-\rho,-\mu)$,
where
$$B(\rho,\mu)=(-\frac{\rho}{2}-\frac{\mu}{k_1}-2q_1p_2+\frac12)_{q_1}+(-\frac{\rho}{2}-\frac{\mu}{k_1}+\frac12)_{q_1}$$
on a generalized eigenbasis.

\subsection{The symmetry $K_0$}
Now we investigate the fact that our method doesn't always give generators of minimal order. In particular, for the case $k_1=k_2=1$ there we have found a
 system of generators of orders $(2,2,2,2,3)$
whereas it is known that a generating set of orders (2,2,2,2,2) exists. In the standard case $k_1=k_2=1$ it is easy to see that ${\cal J}_1$ is of order 2,
 ${\cal K}_1$ is of order 3 and ${\cal K}_2$ is of order 4.  We know that there is a 2nd order symmetry
 operator $K_0$ for this case, independent of $L_2 L_3,H$, such that $[L_3,K_0]=K_1$, $[L_2,K_0]=0$. We will show how to obtain this symmetry from the 
raising and lowering operators
 $K^\pm$, without making use of multiseparability. Thus, for general rational $k_1,k_2$ we look for a symmetry operator $K_0$ such that $[L_3,K_0]=K_1$, $[L_2,K_0]=0$.
 Applying this condition to a formal eigenbasis of functions $\Psi^{n,\rho,m}$  we obtain the general solution
\begin{equation}\label{K0} K_0=-\frac{1}{4p_1q_2}\left(\frac{K^+}{\mu(\mu+p_1p_2)}+\frac{K^-}{\mu(\mu-p_1p_2)}\right)+\xi_m\end{equation}
where $\xi_m$ is a rational scalar function. It is easy to check that the quantity in parentheses is a rational scalar function of $\mu^2$. 
Thus we will have a true constant of the motion, polynomial in the momenta, provided we can choose $\xi_m$ such that the full quantity (\ref{K01})
 is polynomial in $\mu^2$. To determine the possibilities we need to investigate the singularities of this solution  at $\mu=0$ and $\mu=\pm p_1,p_2$, i.e.,
$m=-(b+c+1)/{2}, m=(\pm p_1q_2-b-c-1)/2$. 
Noting that 
$$Y(2)^{-(b+c+1)/2+\ell p_1q_2/(2p_2q_1)}_+=Y(2)^{-(b+c+1)/2-\ell p_1q_2/(2p_2q_1)}_-,\quad X(2)^{-(b+c+1)/2+j}_+=X(2)^{-(b+c+1)/2-j}_-$$
for arbitrary $\ell,j$ we see that $K^+\to K^-$ as $\mu\to 0$ so that the pole at $\mu=0$ is removable. 
 For general  $p_1,p_2,q_1,q_2$ odd we set  
\[\xi_m=-\frac{D_2(L_2)}{2p_1q_2(\mu+p_1p_2)(\mu-p_1p_2)}\]
and determine a polynomial  $D_2$ such that the operator $K_0$ has a removable singularity at $\mu=p_1p_2$, i.e., such that the residue is $0$. (Since the solution is 
a polynomial in $\mu^2$ we don't have to worry about the singularity at $\mu=-p_1p_2$.) 
Thus to compute the residue  we consider the product of operators that forms $K^-$ as applied to basis vectors for which $\mu=p_1p_2$. 
First consider the product of the $p_1q_2$ factors of $X(2)^m_-$:
$$\left(X(2)_-^{(-p_1q_2-b-c)/2}\cdots X(2)^{(-b-c)/2}_-\cdots X(2)^{(p_1q_2-b-c-1)/2}_-\right). $$ 
We make use of the relations 
$$X(2)^{-(b+c+1)/2+j}_+=X(2)^{-(b+c+1)/2-j}_-,\quad X(2)^{m-1}_+X(2)^m_-=4m(m+b)(m+c)(m+b+c).$$
There are a odd number of $X(2)^m_-$ factors in $K^-$ and the central factor is\break $X(2)_-^{-(b+c)/2}=(b^2-c^2)/2$, a constant. The operators on either
 side of this central term  pair up:
\[ X(2)_-^{-(b+c)/2-1} X(2)_-^{-(b+c)/2+1}=X(2)^{-(b+c)/2}_+ X(2)_-^{-(b+c)/2+1}
=4(\frac{b-c+2}{2})(\frac{-b+c+2}{2})(\frac{b+c+2}{2})(\frac{-b-c}{2})\]
which acts as multiplication by a constant. Then we consider the next pair $X(2)_-^{-(b+c)/2-2} X(2)^{-(b+c)/2+2}$, and so on to evaluate the product
$$ \frac{b^2-c^2}{2}\Pi_{m=-(b+c)/2+1}^{m=-(b+c)/2+(p_1q_2-1)/2}\left(4m(m+b)(m+c)(m+b+c)\right).$$ 
Now we consider the $2p_1q_2$  $Y(2)^m_-$ factors in $K^-$: 
$$\left(Y(2)_-^{m_0-(2q_1p_2-1)k_1/2k_2}Y(2)_-^{m_0-(2q_1p_2-2)k_1/2k_2}\cdots
Y(2)_-^{m_0-k_1/2k_2}Y(2)_-^{m_0}\right)$$
where $m_0=(p_1q_2-b-c-1)/2$. We make use of the relations
$$Y(2)^{-(b+c+1)/2+\ell p_1q_2/(2p_2q_1)}_+=Y(2)^{-(b+c+1)/2-\ell p_1q_2/(2p_2q_1)}_-,$$
$$ Y(2)^{m-p_1q_2/(2q_1p_2)}_+Y(2)^m_-=\left(\frac{-\rho+1}{2}-\frac{k_2}{k_1}(2m+b+c+1)\right)\left(\frac{\rho+1}{2}-\frac{k_2}{k_1}(2m+b+c+1)\right)$$
and pair up the 1st and last factors, the 2nd and next to last factors, and so on, to obtain the product
$ \Pi_{\ell=1}^{q_1p_2}((\ell-\frac12)^2-\frac{\rho^2}{4})$. Thus we obtain the residue of the pole and determine that
$$D_2(L_2)=2(c^2-b^2)\Pi_{\ell=1}^{q_1p_2}\left(\ell(\ell-1)+\frac{L_2}{k_1^2}\right)\Pi_{j=1}^{\frac{p_1q_2-1}{2}}(2j-b-c)(2j+b-c)(2j-b+c)(2j+b+c).$$
 The full operator 
\be\label{K01}K_0=-\frac{1}{4p_1q_2}\left(\frac{K^+}{\mu(\mu+p_1p_2)}+\frac{K^-}{\mu(\mu-p_1p_2)}\right)-\frac{D_2(L_2)}{2p_1q_2(\mu+p_1p_2)(\mu-p_1p_2)}\ee
 is thus a polynomial in $\rho^2$ and $\mu^2$, as well as a differential operator in $x,z$, hence it can be defined as a pure differential
 symmetry operator, independent of basis. From the explicit expression for $K_0$ we find easily that
$$ 4p_1q_2K_0(p_1^2p_2^2+L_3)=K_2+p_1p_2K_1+D_2(L_2),$$
which can be written in the  symmetrized form
\be\label{K0symm} 2p_1q_2(2p_1^2p_2^2K_0+\{L_3,K_0\})=K_2+3p_1p_2K_1+D_2(L_2).\ee 
By explicit computation we have checked that this gives the correct 2nd order symmetry operator in the case $k_1=k_2=1$.  Note: A similar construction using the operators $J^\pm$ 
fails to produce a true symmetry operator $J_0$.

Now we consider the symmetry algebra generated by $H, L_2,L_3, K_0, J_1$. 
Using the results of the last two sections we can find algebraic relations between $[J_1,K_0]$ and the generators, so that the symmetry algebra 
closes algebraically. However, it doesn't close polynomially.

\subsection{Two-variable models of the 3-parameter symmetry algebra action}
The recurrence operators introduced via special function theory
lead directly
to two-variable function space models
 representing the symmetry algebra action in terms of difference operators on
spaces of rational functions $f(\rho,\mu)$. Indeed, from relations (\ref{raise1}), (\ref{lower1}), (\ref{raise2}), (\ref{lower2}) it is easy to write down
 a function space model for irreducible
representations of the symmetry algebra. Note that since~$H$ commutes with all
elements of the algebra, it corresponds to multiplication by a constant~$E$ where $E=\alpha^2/u^2$ in
the model. We let complex variables  $\rho,\mu$    correspond to the multiplication realization
\be \label{modeleigenvalues}H=L_1=E,\quad L_2=k_1^2\frac{1-\rho^2}{4},\quad L_3=-\mu^2.\ee
We take a generalized basis function corresponding to eigenvalues $\rho_0,\mu_0$ in the form $\delta(\rho-\rho_0)\delta(\mu-\mu_0)$.
 Then the action of the symmetry 
 algebra on the  space of functions  $f(\rho,\mu)$ is given by
equations
\begin{gather}\label{modelJraising}
J^+f(\rho,\mu)=(\frac{4\alpha}{u})^{p_1}(-\frac{\rho}{2}-\frac{\mu}{k_1}+\frac12)_{q_1}(\frac{u}{2}-\frac{k_1\rho}{2}+\frac12)_{p_1}f(\rho-2q_1,\mu),\\
 \label{modelJlowering}
J^-f(\rho,\mu)=(-\frac{4\alpha}{u})^{p_1}(\frac{\rho}{2}-\frac{\mu}{k_1}+\frac12)_{q_1}
(\frac{u}{2}+\frac{k_1\rho}{2}+\frac12)_{p_1}f(\rho+2q_1,\mu),\\
\label{modelKraising} K^+f(\rho,\mu)=(2)^{p_1q_2}(\frac{\mu}{2k_2}+\frac{b}{2}+\frac{c}{2}+\frac12)_{p_1q_2} (\frac{\mu}{2k_2}+\frac{b}{2}-\frac{c}{2}+\frac12)_{p_1q_2} 
(\frac{1-\rho}{2}+\frac{\mu}{k_1})_{2p_2q_1}f(\rho,\mu+2p_1p_2),\\
\label{modelKlowering} K^-f(\rho,\mu)=(-2)^{p_1q_2}(-\frac{\mu}{2k_2}+\frac{b}{2}+\frac{c}{2}+\frac12)_{p_1q_2} (-\frac{\mu}{2k_2}+\frac{b}{2}-\frac{c}{2}+\frac12)_{p_1q_2} 
(\frac{1-\rho}{2}-\frac{\mu}{k_1})_{2p_2q_1}f(\rho,\mu-2p_1p_2).
 \end{gather}
In terms of  difference operators we have
\begin{gather}\label{modelJraising2}
J^+=(-\frac{4\alpha}{u})^{p_1}(-1)^{q_1}(\frac{\rho}{2}+\frac{\mu}{k_1}+\frac12)_{q_1}(-\frac{u}{2}+\frac{k_1\rho}{2}+\frac12)_{p_1}T_\rho^{2q_1}\\
 \label{modelJlowering2}
J^-=(\frac{4\alpha}{u})^{p_1}(-1)^{q_1}(-\frac{\rho}{2}+\frac{\mu}{k_1}+\frac12)_{q_1}
(-\frac{u}{2}-\frac{k_1\rho}{2}+\frac12)_{p_1}T_\rho^{-2q_1},\\
\label{modelKraising2} K^+=(2)^{p_1q_2}(-\frac{\mu}{2k_2}-\frac{b}{2}-\frac{c}{2}+\frac12)_{p_1q_2} (-\frac{\mu}{2k_2}-\frac{b}{2}+\frac{c}{2}+\frac12)_{p_1q_2} 
(\frac{1+\rho}{2}-\frac{\mu}{k_1})_{2p_2q_1}T_\mu^{-2p_1p_2},\\
\label{modelKlowering2} K^-=(-2)^{p_1q_2}(\frac{\mu}{2k_2}-\frac{b}{2}-\frac{c}{2}+\frac12)_{p_1q_2} (\frac{\mu}{2k_2}-\frac{b}{2}+\frac{c}{2}+\frac12)_{p_1q_2} 
(\frac{1+\rho}{2}+\frac{\mu}{k_1})_{2p_2q_1}T_\mu^{2p_1p_2}.
 \end{gather}
where $T_\rho^A g(\rho,\mu)= g(\rho+A,\mu)$, $T_\mu^B g(\rho,\mu)=g(\rho,\mu+B)$ for rational functions $g(\rho,\mu)$.
Here we have performed a gauge transformation to make the expressions for $J^\pm,K^\pm$ more symmetric. 
The operators $J_1,J_2,K_0,K_1,K_2$ defined in terms of $J^\pm,K^\pm, L_1,L_2,L_3$ by (\ref{J1J2}), (\ref{K1K2}), (\ref{K01}) determine a function space
realization of the symmetry algebra. The space of rational functions $f(\rho,\mu)$ is invariant under the symmetry algebra action, but the space of polynomial functions is not.

\section{The  Kepler-Coulomb system with 4-parameter potential}
Now we study the 4-parameter extended  extended Kepler-Coulomb system (\ref{KC}).
The separation equations for the equations, $H\Psi=E\Psi$,  $L_3\Psi=-\mu^2\Psi$,  $L_2\Psi=\frac{k_1^2}{4}(1-\rho^2)\Psi$ with $\Psi=R(r)\Theta(\theta_1)\Phi(\theta_2)$, are:
$$ 1)\quad (\partial_{\theta_2}^2+\frac{k_2^2(\frac14- b^2)}{\cos^2(k_2\theta_2)}+\frac{k_2^2(\frac14- c^2)}{\sin^2(k_2\theta_2)})\Phi(\theta_2)=-\mu^2\Phi(\theta_2),$$
where we have taken $\beta =k_2^2(\frac14- b^2), \gamma= k_2^2(\frac14- c^2)$, and we write $\mu=k_2(2m+b+c+1)$;
$$ 2)\quad (\partial_{\theta_1}^2 +k_1\cot(k_1\theta_1)\partial_{\theta_1}-\frac{\mu^2}{\sin^2(k_1\theta_1)}+\frac{k_2^2(\frac14-d^2)}{\cos^2(k_1\theta_1)})\Theta(\theta_1)
=\frac{k_1^2}{4}(1-\rho^2)\Theta(\theta_1),$$ 
where we have taken  $\delta=k_1^2(\frac14-d^2)$.
If we look for solutions of the form 
$$\Theta(\theta_1)=\frac{\Psi(\theta_1)}{\sqrt{\sin(k_1\theta_1)}}$$
we see that the differential equation satisfied by $\Psi$ is
$$2)'\quad (\partial_{\theta_1}^2+\frac{\frac{k_1^2}{4}-\mu^2}{\sin^2(k_1\theta_1)}+\frac{k_1^2(\frac14-d^2)}{\cos^2(k_1\theta_1)}+\frac{k_1^2\rho^2}{4})\Psi(\theta_1)=0.$$
Further
$$ 3)\quad (\partial _r^2+\frac{2}{r}\partial_r + \frac{1-k_1^2\rho^2}{4r^2}+\frac{\alpha}{r}-E)R(r)=0.$$
In 3) we introduce a new variable ${\rm R}=2\sqrt{E}r$ so that the separation equation becomes
$$ 3)' \quad (\partial _{{\rm R}}^2 + \frac{1-k_1^2\rho^2}{4{\rm R}^2}+\frac{\alpha}{2\sqrt{E}{\rm R}}-\frac14)S({\rm R})=0,$$
with $R(r)=S({\rm R})/{\rm R}$.
The separated solutions are 
$$\Xi_{n,m,p}=R^{k_1\rho}_n(r)\Phi_m^{(c,b)}(\cos(2k_2\theta_2)) \frac{\Psi^{(\mu/k_1,d)}_p(\cos k_1\theta_1)}{\sqrt{\sin(k_1\theta_1)}},$$
$$\Phi_m^{(c,b)}(\cos(2k_2\theta_2))=\sin^{c+1/2}(k_2\theta_2)\cos^{b+1/2}(k_2\theta_2)P_m^{(c,b)}(\cos(2k_2\theta_2)),$$
$$ \Psi^{(\mu/k_1,d)}_p(\cos(k_1\theta_1))=\sin^{\mu/k_1+1/2}(k_1\theta_1)\cos^{d+1/2}(k_1\theta_1)P^{(\mu/k_1,d)}_p(\cos(2k_1\theta_1)),$$
where the $P_m^{(c,b)}(\cos(2k_2\theta_2))$, $P^{(\mu/k_1,d)}_p(\cos(2k_1\theta_1))$ are Jacobi  functions; 
$$ R^{k_1\rho}_n(r)=\frac{S(r)}{2\sqrt{E}r},\quad S(r)=(2\sqrt{E})^{k_1\rho/2}e^{-\sqrt{E}r}r^{(k_1\rho+1)/2}L_n^{k_1\rho}(2\sqrt{E}r),$$
and $\rho=2(2p+\frac{\mu}{k_1}+d+1)$, where the $L_n^{k_1\rho}(2\sqrt{E}r)$ are associated Laguerre functions, and the relation 
between $E$, $\rho$ and $n$ is the quantization condition (where we set $u=2n+k_1\rho+1$)
\be\label{E_n1} E=\frac{\alpha^2}{u^2}=\frac{\alpha^2}{(2[n+2k_1p+2k_2m]+2k_1[d+1]+2k_2[b+c+1]+1)^2}.\ee

 As in the computations with the 3-parameter potential, we  are  interested in the space of generalized eigenfunctions, 
not the normalization of any individual eigenfunction. Thus our relations  are valid on generalized eigenspaces and don't 
 agree with the normalization of usual polynomial eigenfunctions.

There are transformations that preserve $E$ and imply quantum superintegrability. Indeed for $k_1=p_1/q_1, k_2=p_2/q_2$ the following transformations 
will accomplish this:
\[ 1):\ n\to n+2p_1,\quad m\to m,\quad p\to p-q_1,\
 2):\  n\to n-2p_1,\quad m\to m,\quad p\to p+q_1,\]
\[3):\ n\to n,\quad m\to m-p_1q_2, \quad p\to p+q_1p_2,\
4):\ n\to n,\quad m\to m+p_1q_2, \quad p\to p-q_1p_2.\]

To effect the $r$-dependent transformations 1), 2) we use $Y(1)^{\pm n}$  as in(\ref{Dnpmt}).
To  incorporate the $\theta_1$-dependent parts of 1), 2)  we use recurrence formulas for the
functions $\Psi_p^{\mu/k_1,d}(z)$ where $z=\cos(2k_1\theta_1)$:
\[ Z(1)^p_-\frac{\Psi_p^{\mu/k_1,d}(z)}{{\sqrt{\sin(k_1\theta_1)}}}\equiv
 \left(-(1-z^2)(1-\frac{\rho}{2})\partial_z+\frac12((1-\frac{\rho}{2})(-\frac{\rho}{2})z+
\frac{\mu^2}{k_1^2}-d^2)\right)\frac{\Psi_p^{\mu/k_1,d}(z)}{{\sqrt{\sin(k_1\theta_1)}}}\]
\[=-2(\frac{\rho}{4}+\frac{\mu}{2k_1}-\frac{d}{2}-\frac12)(\frac{\rho}{4}-\frac{\mu}{2k_1}+
 \frac{d}{2}-\frac12)\frac{\Psi_{p-1}^{\mu/k_1,d}(z)}{{\sqrt{\sin(k_1\theta_1)}}},\]
\[ Z(1)^p_+\frac{\Psi_p^{\mu/k_1,d}(z)}{{\sqrt{\sin(k_1\theta_1)}}}\equiv
 \left(-(1-z^2)(1+\frac{\rho}{2})\partial_z+\frac12((1+\frac{\rho}{2})(\frac{\rho}{2})z+
\frac{\mu^2}{k_1^2}-d^2)\right)\frac{\Psi_p^{\mu/k_1,d}(z)}{{\sqrt{\sin(k_1\theta_1)}}}\]
\[=-2(\frac{\rho}{4}+\frac{\mu}{2k_1}+\frac{d}{2}+\frac12)(\frac{\rho}{4}-\frac{\mu}{2k_1}-
 \frac{d}{2}+\frac12)\frac{\Psi_{p+1}^{\mu/k_1,d}(z)}{{\sqrt{\sin(k_1\theta_1)}}},\]
We see that the operators $Z(1)_\pm^p$ depend 
on $\mu^2$ (which can be interpreted as a differential operator) and are polynomial  in $\rho$. We now
form the operators 
$$J^+ =\left(Y(1)^{n+2p_1-1}_+Y(1)^{n+2p_1-2}_+\cdots Y(1)^{n+1}_+Y(1)^n_+\right)\left(Z(1)^{p-(q_1-1)}_-\cdots Z(1)^p_-\right),$$ 
$$J^- =\left(Y(1)^{n-2p_1+1}_-Y(1)^{n-2p_1+2}_-\cdots Y(1)^{n-1}_- Y(1)^n_-\right)\left(Z(1)^{p+(q_1-1)}_+\cdots Z(1)^p_+\right),$$ 
Since $J^+$ and $J^-$ switch places under the reflection $\rho\to -\rho$  we see that 
$$J_2=J^+ +J^-,\quad J_1=(J^--J^+)/\rho$$
 are even functions in
 both $\rho$ and $\mu$  and can be interpreted as  differential operators. 
 
 To implement the $\theta_1$-dependent parts of 3),  4) we set $w=\sin^2(k_1\theta_1)$ and use 
 $$W^p_-(1)\frac{\Psi_p^{\mu/k_1,d}}{w^{1/4}}\equiv 
\left[(1+\frac{\mu}{k_1})(w-1)\frac{d}{dw}-\frac{\mu}{4k_1}(1-\frac{2}{w})(1+\frac{\mu}{k_1})+\frac{d^2-\rho^2/4}{4}\right]
\frac{\Psi_p^{\mu/k_1,d}}{w^{1/4}}$$ $$=\frac{(\frac{\mu}{k_1}+\frac{\rho}{2}+d+1)(-\frac{\mu}{k_1}+\frac{\rho}{2}+d-1)}{4(\frac{\mu}{k_1}+1)
(\frac{\mu}{k_1}+2)}\frac{\Psi_{p-1}^{\mu/k_1+2,d}}{w^{1/4}},$$
 $$W^p_+(1)\frac{\Psi_p^{\mu/k_1,d}}{w^{1/4}}\equiv \left[(1-\frac{\mu}{k_1})(w-1)
\frac{d}{dw}+\frac{\mu}{4k_1}(1-\frac{2}{w})(1-\frac{\mu}{k_1})+\frac{d^2-\rho^2/4}{4}\right]\frac{\Psi_p^{\mu/k_1,d}}{w^{1/4}}$$ 
$$=\frac14(-\frac{\mu}{k_1}+\frac{\rho}{2}-d+1)(\frac{\mu}{k_1}+\frac{\rho}{2}-d-1)\frac{\mu}{k_1}(\frac{\mu}{k_1}-1)\frac{\Psi_{p+1}^{\mu/k_1-2,d}}{w^{1/4}},$$
 To implement the $\theta_2$-dependent parts of 3) and 4) we use the recurrences $X(2)^m_\pm$ already employed for the 3-parameter potential, (\ref{X2}).

We define 
$$K^+ =\left(W(1)^{p+(q_1p_2-1)}_+\cdots W(1)^p_+\right)\left(X(2)^{m-(p_1q_2-1)}_-\cdots X(2)^m_-\right),$$ 
$$K^- =\left(W(1)^{p-(q_1p_2-1)}_-\cdots W(1)^p_-\right)\left(X(2)^{m+(p_1q_2-1)}_+\cdots X(2)^m_+\right).$$ 
From the form of these operators we see that they are even functions of $\rho^2$ and they switch places under the reflection $\mu\to -\mu$. Thus
$ K_2= K^++K^-$, $ K_1=(K^--K^+)/\mu$
 are even functions in
 both $\rho$ and $\mu$  and can be interpreted as  differential operators.

 We have the action of $J^\pm,K^\pm$ on a generalized eigenbasis:
 \bea\label{Jpm2} J^+\Xi_{n,m,p}&=&\frac{(-\frac{\rho}{4}-\frac{\mu}{2k_1}+\frac{d}{2}+\frac12)_{q_1}(-\frac{\rho}{4}+\frac{\mu}{2k_1}-
\frac{d}{2}+\frac12)_{q_1}(\frac{v}{2}-\frac{k_1\rho}{2}+\frac12)_{2p_1}(-\frac{v}{2}-\frac{k_1\rho}{2}-\frac12)_{2p_1}}
{(2)^{-4p_1-q_1}(-1)^{q_1}\alpha^{-2p_1}u^{2p_1}}\Xi_{n+2p_1,m,p-q_1},\nonumber\\
J^-\Xi_{n,m,p}&=&\frac{(\frac{\rho}{4}-\frac{\mu}{2k_1}-\frac{d}{2}+\frac12)_{q_1}(\frac{\rho}{4}+\frac{\mu}{2k_1}+\frac{d}{2}+\frac12)_{q_1}}
{(2)^{-4p_1-q_1}(-1)^{q_1}\alpha^{-2p_1}u^{2p_1}}\Xi_{n-2p_1,m,p+q_1},\eea
\be\label{Kpm2} K^+\Xi_{n,m,p}=(-1)^{q_1p_2}2^{p_1q_2}(-\frac{\mu}{k_1})_{2q_1p_2}\times\ee
\[(\frac{\rho}{4}-\frac{\mu}{2k_1}-\frac{d}{2}+\frac12)_{q_1p_2}(-\frac{\rho}{4}-\frac{\mu}{2k_1}+\frac{d}{2}+\frac12)_{q_1p_2}
(-\frac{\mu}{2k_2}+\frac{b}{2}-\frac{c}{2}+\frac12)_{p_1q_2}
(-\frac{\mu}{2k_2}-\frac{b}{2}+\frac{c}{2}+\frac12)_{p_1q_2}\Xi_{n,m-p_1q_2,p+q_1p_2},\]
$$K^-\Xi_{n,m,p}=(-1)^{q_1p_2}2^{p_1q_2}\times$$
$$\frac{(\frac{\rho}{4}+\frac{\mu}{2k_1}+\frac{d}{2}+\frac12)_{q_1p_2}(-\frac{\rho}{4}+\frac{\mu}{2k_1}-\frac{d}{2}+\frac12)_{q_1p_2}
(\frac{\mu}{2k_2}-\frac{b}{2}-\frac{c}{2}+\frac12)_{p_1q_2}(\frac{\mu}{2k_2}+\frac{b}{2}+\frac{c}{2}+\frac12)_{p_1q_2}}{(\frac{\mu}{k_1}+1)_{2q_1p_2}}\Xi_{n,m+p_1q_2,p-q_1p_2}.$$
Further, we recall that 
\[ H\Xi_{n,m,p}=\frac{\alpha^2}{(2n+k_1\rho+1)^2}\Xi_{n,m,p}=E\Xi_{n,m,p},\ L_2\Xi_{n,m,p}=\frac{k_1^2}{4}(1-\rho^2)\Xi_{n,m,p},\
L_3\Xi_{n,m,p}=-\mu^2\Xi_{n,m,p},\]
\[ u=2n+k_1\rho+1=\frac{\alpha}{\sqrt{E}},\quad \rho=2(2p+\mu/k_1+d+1),\quad \mu=k_2(2m+b+c+1).\]

Using these 
definitions  and computing on an eigenbasis, it is straightforward to verify the relations: 
\[[J_1,L_2]=2k_1^2q_1J_2+4p_1^2J_1,\
 [J_2,L_2]=-4q_1\{J_1,L_2\}-4q_1^2k_1^2J_2+2k_1^2q_1(1-8q_1^2)J_1,\]
\[[J_1,L_3]=[J_2,L_3]=0.\
[K_1,L_3]=4p_1p_2K_2+4p_1^2p_2^2K_1,\]
\[[K_2,L_3]=-2p_1p_2\{L_3,K_1\}-4p_1^2p_2^2K_2-8p_1^3p_2^3K_1,\
[K_1,L_2]=[K_2,L_2]=0.\]

Further we find 
$$J^+J^-\Xi_{n,m,p}=$$
$$4^{4p_1+q_1}E^{2p_1}(\frac{\rho/2-\frac{\mu}{k_1}-d+1}{2})_{q_1}(\frac{\rho/2+\frac{\mu}{k_1}+d+1}{2})_{q_1}(\frac{\rho/2+\frac{\mu}{k_1}-d+1}{2})_{q_1}
(\frac{\rho/2-\frac{\mu}{k_1}+d+1}{2})_{q_1}\times$$
$$(\frac{k_1\rho-\frac{\alpha}{\sqrt{E}}+1}{2})_{2p_1}(\frac{k_1\rho+\frac{\alpha}{\sqrt{E}}+1}{2})_{2p_1}\Xi_{n,m,p}$$
$$J^-J^+\Xi_{n,m,p}=$$
$$4^{4p_1+q_1}E^{2p_1}(\frac{-\rho/2-\frac{\mu}{k_1}-d+1}{2})_{q_1}(\frac{-\rho/2+\frac{\mu}{k_1}+d+1}{2})_{q_1}(\frac{-\rho/2+\frac{\mu}{k_1}-d+1}{2})_{q_1}
(\frac{-\rho/2-\frac{\mu}{k_1}+d+1}{2})_{q_1}\times$$
$$(\frac{-k_1\rho-\frac{\alpha}{\sqrt{E}}+1}{2})_{2p_1}(\frac{-k_1\rho+\frac{\alpha}{\sqrt{E}}+1}{2})_{2p_1}\Xi_{n,m,p}$$
$$K^+K^-\Xi_{n,m,p}=$$
$$2^{2p_1q_2}(\frac{\rho/2+\frac{\mu}{k_1}+d+1}{2})_{q_1p_2}(\frac{\rho/2+\frac{\mu}{k_1}-d+1}{2})_{q_1p_2}(\frac{-\rho/2+\frac{\mu}{k_1}-d+1}{2})_{q_1p_2}
(\frac{-\rho/2+\frac{\mu}{k_1}+d+1}{2})_{q_1p_2}\times$$
$$(\frac{\mu}{2k_2}-\frac{b}{2}-\frac{c}{2}+\frac12)_{p_1q_2}(\frac{\mu}{2k_2}+\frac{b}{2}+\frac{c}{2}+\frac12)_{p_1q_2}
(\frac{\mu}{2k_2}-\frac{b}{2}+\frac{c}{2}+\frac12)_{p_1q_2}(\frac{\mu}{2k_2}+\frac{b}{2}-\frac{c}{2}+\frac12)_{p_1q_2}
\Xi_{n,m,p},$$
 $$K^-K^+\Xi_{n.m.p}=$$ $$2^{2p_1q_2}(\frac{\rho/2-\frac{\mu}{k_1}+d+1}{2})_{q_1p_2}(\frac{\rho/2-\frac{\mu}{k_1}-d+1}{2})_{q_1p_2}
(\frac{-\rho/2-\frac{\mu}{k_1}-d+1}{2})_{q_1p_2}(\frac{-\rho/2-\frac{\mu}{k_1}+d+1}{2})_{q_1p_2}\times$$
 $$(\frac{-\mu}{2k_2}-\frac{b}{2}-\frac{c}{2}+\frac12)_{p_1q_2}(\frac{-\mu}{2k_2}+\frac{b}{2}+\frac{c}{2}+\frac12)_{p_1q_2}
(\frac{-\mu}{2k_2}-\frac{b}{2}+\frac{c}{2}+\frac12)_{p_1q_2}(\frac{-\mu}{2k_2}+\frac{b}{2}-\frac{c}{2}+\frac12)_{p_1q_2}\Xi_{n,m,p}.$$
 From these expressions it is easy to see that each of $J^+J^-,J^-J^+$ is a polynomial in $\mu^2$ and $E$ and that these operators switch places 
under the reflection $\rho\to-\rho$. Thus $P_1(H,L_2,L_3)=J^+J^-+J^-J^+$ and $P_2(H,L_2,L_3)=(J^+J^--J^-J^+)/\rho$ are each polynomials in $H,L_2,L_3$.
 Similarly, each of $K^+K^-,K^-K^+$ is a polynomial in $\rho^2$ and in $\mu$ and  they switch places under the reflection $\mu\to -\mu$. 
Thus $P_3(L_2,L_3)=K^+K^-+K^-K^+$, $P_4(L_2,L_3)=(K^+K^--K^-K^+)/\mu$ are polynomials in $L_2,L_3$.
 
 Straightforward consequences of these formulas are the structure relations
 $$[J_1,J_2]=-2q_1J_1^2-2P_2(H,L_2,L_3),\quad [K_1,K_2]=-2p_1p_2K_1^2-2P_4(L_2,L_3),$$
 and the unsymmetrized structure relations
 $$J_1^2(\frac14-k_1^{-2}L_2)=J_2^2-2P_1(H,L_2,L_3)-2q_1J_1J_2,\quad K_1^2=-K_2^2+2P_3(L_2,L_3)+2p_1p_2K_1K_2.$$

\subsection{Lowering the orders of the generators}
Just as for the classical analogs, we can find generators that are of order one less than $J_1$ and $K_1$. First we look for a symmetry operator $J_0$ such 
that $[L_2,J_0]=J_1$ and $[L_3,J_0]=0$:
$$J_0=-\frac{1}{2k_1^2q_1}\left(\frac{J^-}{\rho(\rho+2q_1)}+\frac{J^+}{\rho(\rho-2q_1)}\right)+\frac{S_1(H,L_3)}{\rho^2-4q_1^2},$$
where the symmetry operator $S_1$ is to be determined. From this it is easy to show that 
$$2k_1^2q_1(\rho^2-4q_1^2)J_0=-J_2+2q_1J_1+2k_1^2q_1S_1(H,L_3).$$
To determine $S_1$ we evaluate both sides of this equation for $\rho=-2q_1$:
$S_1(H, L_3)=\frac{1}{k_1^2q_1}J^-_{\rho=-2q_1}$,
and apply $J^-$ to a specific eigenfunction.
Consider the product of the $2p_1$ factors of $Y(1)^n_-$:
$$Y(1)_-^{n_0-(2p_1-1)}Y(1)^{n_0-(2p_1-2)}_-\cdots  Y(1)^{n_0-1}_-Y(1)^{n_0}_- $$ 
 acting on an eigenbasis function corresponding to  $n_0=p_1+(\alpha/\sqrt{E}-1)/2$.
There are an even number of $Y(1)^\ell_-$ factors in $J^-$ and the central pair is $Y(1)^{n_0-p_1}_-Y(1)^{n_0-(p_1-1)}_-$.  
The left hand operator corresponds to $k_1\rho=0$ and the right hand operator corresponds to $k_1\rho=-2$. Thus we have 
$$Y(1)^{n_0-p_1}_-Y(1)^{n_0-(p_1-1)}_-=\left(2\partial_r+(2\alpha+\frac{1}{r}\right)\left(-2\partial_r+(2\alpha-\frac{3}{r}\right)=Y(1)^{n_0-p_1}_+Y(1)^{n_0-(p_1-1)}_-
=4(\alpha^2-E).$$
Operators on either side of the central term can be grouped in nested pairs $Y(1)^{n_0-p_1-s}_-Y(1)^{n_0-(p_1-1)+s}_-$, where $s=0,1,\cdots,p_1-1$, 
corresponding to $k_1\rho=2s$ $k_1\rho=-2(s+1)$, respectfully.  Each pair gives
$$Y(1)^{n_0-p_1-s}_-Y(1)^{n_0-(p_1-1)+s}_-=\left(2(2s+1)\partial_r+(2\alpha+\frac{1-4s^2}{r}\right)\left(2(-2s-1)\partial_r+(2\alpha+\frac{1-4(s+1)^2}{r}\right)$$
$$=Y(1)^{n_0-p_1+s}_+Y(1)^{n_0-(p_1-1)+s}_-
=4(\alpha^2-(1+2s)^2E).$$
Thus,
\[Y(1)_-^{n_0-(2p_1-1)} Y(1)^{n_0-(2p_1-2)}_-\cdots Y(1)^{n_0}_- 
=4^{p_1} \Pi_{s=0}^{p_1-1}\left(\alpha^2-(1+2s)^2E\right).\] 

Now we consider the $q_1$ factors  $Z(1)^\ell_+$  in $J^-$:
$$Z(1)_+^{p_0+(q_1-1)}Z(1)_+^{p_0+(q_1-2)}\cdots
Z(1)_+^{p_0}$$
acting on an eigenbasis function corresponding to  $p_0=-(q_1+\mu/k_1+d+1)/2$. There are an odd number of factors and the central factor is $(\mu^2/k_1^2-d^2)/2$. 
Operators on either side of this central term can be grouped in nested pairs to give
$$Z(1)_+^{p_0+(q_1-1)}Z(1)_+^{p_0+(q_1-2)}\cdots
Z(1)_+^{p_0}=-2(\frac{\mu^2}{k_1^2}-d^2)(-4)^{(q_1-1)/2}\Pi_{s=1}^{(q_1-1)/2}((s-\frac{d}{2})^2-\frac{\mu^2}{4k_1^2})((s+\frac{d}{2})^2-\frac{\mu^2}{4k_1^2}),$$ 
$$S_1(H,L_3)=$$
$$\frac{1}{2k_1^2q_1}(\frac{L_3}{k_1^2}+d^2)(-4)^{(q_1-1)/2}4^{p_1}\ \Pi_{s=0}^{p_1-1}\left(\alpha^2-(1+2s)^2H\right)\Pi_{s=1}^{(q_1-1)/2}((s-\frac{d}{2})^2+\frac{L_3}{4k_1^2})((s+\frac{d}{2})^2+\frac{L_3}{4k_1^2}),$$
$$[L_2,J_0]=J_1,\quad [L_3,J_0]=0,\quad 2k_1^2q_1J_0\cdot(1-\frac{4L_2}{k_1^2}-4q_1^2)=-J_2+2q_1J_1+2k_1^2q_1S_1(H,L_3).$$

Next we look for a symmetry operator $K_0$ such that $[L_3,K_0]=K_1$ and $[L_2,K_0]=0$: 
$$K_0=-\frac{1}{4p_1p_2}\left(\frac{K^-}{\mu(\mu+p_1p_2)}+\frac{K^+}{\mu(\mu-p_1p_2)}\right)+\frac{S_2(L_2)}{\mu^2-p_1^2p_2^2},$$
where the symmetry operator $S_2$ is to be determined. From this it is easy to show that 
$$4p_1p_2K_0(\mu^2-p_1^2p_2^2)=-K_2+p_1p_2K_1+4p_1p_2S_2(L_2).$$
To determine $S_2$ we evaluate both sides of this equation for $\mu=-p_1p_2$:
$S_2(L_2)=\frac{1}{2p_1p_2}K^-_{\mu=-p_1p_2}$ and apply $K^-$ to a specific eigenfunction.
 Consider the product of the $q_1p_2$ factors of $W(1)^p_-$:
$$\left(W(1)_-^{(p_0-(q_1p_2-1)}\cdots W(1)^{p_0-(q_1p_2-2)}_-\cdots W(1)^{p_0}_-\right) $$ 
where $p_0=(\rho-q_1p_2-d-1)/2$.
There are a odd number of $W(1)^\ell_-$ factors in $K^-$ and the central factor is $(d^2-\rho^2/4)/4$. Operators on either side of 
this central term can be grouped in nested pairs:
\[\left(W(1)_+^{(p_0+(q_1p_2-1)}\cdots W(1)^{p_0+(q_1p_2-2)}_+\cdots W(1)^{p_0}_+\right)
=(\frac{d^2}{4}-\frac{\rho^2}{16})\Pi_{s=1}^{(q_1p_2-1)/2}\left((\frac{d}{2}+\frac{\rho}{4})^2-s^2\right)\left((\frac{d}{2}-\frac{\rho}{4})^2-s^2\right).\]

Now we consider the $p_1q_2$ factors  $X(2)^\ell_+$  in $K^-$:
$$\left(X(2)_+^{m_0+(p_1q_2-1)}X(2)_+^{m_0+(p_1q_2-2)}\cdots
X(2)_+^{m_0}\right)$$
where $m_0=-(p_1q_2+b+c+1)/2$. There are an odd number of factors and the central factor is $(b^2-c^2)/2$. The operators on either side of this central term 
can be grouped in nested pairs to give:
$$\left(X(2)_+^{m_0+(p_1q_2-1)}X(2)_+^{m_0+(p_1q_2-2)}\cdots
X(2)_+^{m_0}\right)=$$
$$2^{p_1q_2-2}(b^2-c^2)\Pi_{s=1}^{(p_1q_2-1)/2}(s+\frac{c+b}{2})(s-\frac{c+b}{2})(s-1-\frac{b-c}{2})(s-1+\frac{b-c}{2}),$$
\be\label{S2L2} S_2(L_2)=(d^2-\frac{\rho^2}{4})2^{p_1q_2-5}\frac{(b^2-c^2)}{p_1p_2}\Pi_{s=1}^{(q_1p_2-1)/2}\left((\frac{d}{2}+\frac{\rho}{4})^2-s^2\right)\left((\frac{d}{2}-\frac{\rho}{4})^2-s^2\right)\times\ee
$$\Pi_{s=1}^{(p_1q_2-1)/2}(s+\frac{c+b}{2})(s-\frac{c+b}{2})(s-1-\frac{b-c}{2})(s-1+\frac{b-c}{2}).$$
We conclude that
$$ [L_3,K_0]=K_1,\quad[L_2,K_0]=0,\quad 4p_1p_2K_0(L_3+p_1^2p_2^2)=K_2-p_1p_2K_1-4p_1p_2S_2(L_2).$$

\subsection{The structure equations}

Now we determine the commutators of the $J$-operators with the $K$-operators.
We write 
\[ J^+\Xi_{n,m,p}={\cal J}^+(n,m,p)\ \Xi_{n+2p_1,m,p-q_1},\
 J^-\Xi_{n,m,p}={\cal J}^-(n,m,p)\ \Xi_{n-2p_1,m,p+q_1},\]
\[ K^+\Xi_{n,m,p}={\cal K}^+(m,p)\ \Xi_{n,m-p_1q_2,p+q_1p_2},\
 K^-\Xi_{n,m,p}={\cal K}^-(m,p)\ \Xi_{n,m+p_1q_2,p-q_1p_2},\]
where ${\cal J}^\pm,{\cal K}^\pm$ are defined by the right hand sides of equations (\ref{Jpm2}),  (\ref{Kpm2}).
Then, we find
\[[J^+,K^+]\Xi_{n,m,p}=\left(1-A(-\rho,\mu,)\right)J^+K^+\Xi_{n,m,p},\
[J^-,K^+]\Xi_{n,m,p}=\left(1-A(\rho,\mu)\right)J^-K^+\Xi_{n,m,p},\]
\[[J^+,K^-]\Xi_{n,m,p}=\left(1-A(-\rho,-\mu)\right)J^+K^-\Xi_{n,m,p},\
[J^-,K^-]\Xi_{n,m,p}=\left(1-A(\rho,-\mu)\right)J^-K^-\Xi_{n,m,p},\]
or
$$[J^+,K^+]\Xi_{n,m,p}=\frac{1-A(-\rho,\mu,)}{1+A(-\rho,\mu,)}\{J^+,K^+\}\Xi_{n,m,p}\equiv C(-\rho,\mu)\{J^+,K^+\}\Xi_{n,m,p},$$
$$[J^-,K^+]\Xi_{n,m,p}=\frac{1-A(\rho,\mu)}{1+A(\rho,\mu)}\{J^-,K^+\}\Xi_{n,m,p}\equiv C(\rho,\mu)\{J^-,K^+\}\Xi_{n,m,p},$$
$$[J^+,K^-]\Xi_{n,m,p}=\frac{1-A(-\rho,-\mu)}{1+A(-\rho,-\mu)}\{J^+,K^-\}\Xi_{n,m,p}\equiv C(-\rho,-\mu)\{J^+,K^-\}\Xi_{n,m,p},$$
$$[J^-,K^-]\Xi_{n,m,p}=\frac{1-A(\rho,-\mu)}{1+A(\rho,-\mu)}\{J^-,K^-\}\Xi_{n,m,p}\equiv C(\rho,-\mu)\{J^-,K^-\}\Xi_{n,m,p},$$
$$A(\rho,\mu)=
\frac{(\frac{\rho}{4}+\frac{\mu}{2k_1}+\frac12-\frac{d}{2})_{q_1}(\frac{\rho}{4}+\frac{\mu}{2k_1}+\frac12+\frac{d}{2})_{q_1}}
{(\frac{\rho}{4}+\frac{\mu}{2k_1}+\frac12-q_1p_2-\frac{d}{2})_{q_1}(\frac{\rho}{4}+\frac{\mu}{2k_1}+\frac12-q_1p_2+\frac{d}{2})_{q_1}}.$$
From this we can compute the relations
$$[J_1,K_1]=Q^{11}_{11}\{J_1,K_1\}+Q^{11}_{12}\{J_1,K_2\}+Q^{11}_{21}\{J_2,K_1\}+Q^{11}_{22}\{J_2,K_2\},$$
$$[J_2,K_2]=Q^{22}_{11}\{J_1,K_1\}+Q^{22}_{12}\{J_1,K_2\}+Q^{22}_{21}\{J_2,K_1\}+Q^{22}_{22}\{J_2,K_2\},$$
where the $Q^{hh}_{j\ell}$ are rational functions of $\rho^2$ and $\mu^2$. In particular,
$$ Q^{11}_{11}= Q^{22}_{22}=\frac14\left(C(-\rho,\mu)+C(\rho,\mu)+C(-\rho,-\mu)+C(\rho,-\mu)\right),$$
$$Q^{11}_{12}=\frac{1}{\mu^2}Q^{22}_{21}=\frac{1}{4\mu}\left(-C(-\rho,\mu)-C(\rho,\mu)+C(-\rho,-\mu)+C(\rho,-\mu)\right),$$
$$Q^{11}_{21}=\frac{1}{\rho^2}Q^{22}_{12}=\frac{1}{4\rho}\left(-C(-\rho,\mu)+C(\rho,\mu)-C(-\rho,-\mu)+C(\rho,-\mu)\right),$$
$$ Q^{11}_{22}=\frac{1}{\mu^2\rho^2}Q^{22}_{11}=\frac{1}{4\mu\rho}\left(C(-\rho,\mu)-C(\rho,\mu)-C(-\rho,-\mu)+C(\rho,-\mu)\right),$$
Similarly we find
$$[J_1,K_2]=Q^{12}_{11}\{J_1,K_1\}+Q^{12}_{12}\{J_1,K_2\}+Q^{12}_{21}\{J_2,K_1\}+Q^{12}_{22}\{J_2,K_2\},$$
$$[J_2,K_1]=Q^{21}_{11}\{J_1,K_1\}+Q^{21}_{12}\{J_1,K_2\}+Q^{21}_{21}\{J_2,K_1\}+Q^{21}_{22}\{J_2,K_2\},$$
\[ Q^{12}_{11}=Q^{22}_{21},\ Q^{12}_{12}=Q^{21}_{21}=Q^{11}_{11},\ Q^{21}_{11}=Q^{22}_{12},\ Q^{12}_{21}=\mu^2 Q^{11}_{22},\
Q^{12}_{22}=\frac{1}{\rho^2} Q^{22}_{12},\ Q^{21}_{12}=\frac{1}{\mu^2}Q^{22}_{11},\ 
 Q^{21}_{22}=\frac{1}{\mu^2}Q^{22}_{21}.\]
These relations make sense only on a generalized eigenbasis, but can be cast into  operator form 
\be\label{ratstructure}[K_\ell, J_h] Q=\{J_1,K_1\}P^{h\ell}_{11}+\{J_1,K_2\}P^{h\ell}_{12}+\{J_2,K_1\}P^{h\ell}_{21}+\{J_2,K_2\}P^{h\ell}_{22},\quad h,\ell=1,2,\ee
where $Q$ and $P^{h\ell}_{jk}$ are polynomial operators in $L_2,L_3$. In particular, $Q$ is the symmetry operator defined by the product
$$B(\rho,\mu)B(-\rho,\mu)B(\rho,-\mu)B(-\rho,-\mu),$$
$$
B(\rho,\mu)=(\frac{\rho}{4}+\frac{\mu}{2k_1}+\frac12-\frac{d}{2})_{q_1}(\frac{\rho}{4}+\frac{\mu}{2k_1}+\frac12+\frac{d}{2})_{q_1}+(\frac{\rho}{4}+\frac{\mu}{2k_1}+\frac12-q_1p_2-\frac{d}{2})_{q_1}(\frac{\rho}{4}+\frac{\mu}{2k_1}+\frac12-q_1p_2+\frac{d}{2})_{q_1} $$
on a generalized eigenbasis.
Thus for general $k_1,k_2$ the symmetry algebra closes only algebraically.The basis generators are $H,L_2,L_3,J_0,K_0$ and the commutators $J_1,K_1$  are appended to the algebra.

\subsection{Two-variable models of the 4-parameter symmetry algebra action}
Our recurrence operators 
lead directly
to two-variable function space models
that represent the symmetry algebra action in terms of difference operators on
spaces of rational functions $f(\rho,\mu_)$. Using relations (\ref{Jpm2}), (\ref{Kpm2}) we can determine a function space model for irreducible
representations of the symmetry algebra. Since~$H$ commutes with all
elements of the algebra, it corresponds to multiplication by   $E=\alpha^2/u^2$ in
the model. We let complex variables  $\rho,\mu$    correspond to the  realization
\be \label{modeleigenvalues1}H=L_1=E,\quad L_2=k_1^2\frac{1-\rho^2}{4},\quad L_3=-\mu^2.\ee
We take a generalized basis function corresponding to eigenvalues $\rho_0,\mu_0$ in the form $\delta(\rho-\rho_0)\delta(\mu-\mu_0)$.
 Then the action of the  symmetry 
 algebra on the  space of functions  $f(\rho,\mu)$ is given by 
\be\label{modelJraising1}
J^+f(\rho,\mu)=\ee
\[\frac{\alpha^{2p_1}(-\frac{\rho}{4}-\frac{\mu}{2k_1}+\frac{d}{2}+\frac12)_{q_1}(-\frac{\rho}{4}+\frac{\mu}{2k_1}-\frac{d}{2}+\frac12)_{q_1}(\frac{u}{2}-\frac{k_1\rho}{2}+\frac12)_{2p_1}(-\frac{u}{2}-\frac{k_1\rho}{2}+\frac12)_{2p_1}}{(2)^{-4p_1-q_1}(-1)^{-q_1}u^{2p_1}}f(\rho-4q_1,\mu),\]
\be \label{modelJlowering1}
J^-f(\rho,\mu)=
\frac{\alpha^{2p_1}(\frac{\rho}{4}-\frac{\mu}{2k_1}-\frac{d}{2}+\frac12)_{q_1}(\frac{\rho}{4}+\frac{\mu}{2k_1}+\frac{d}{2}+\frac12)_{q_1}}{(2)^{-4p_1-q_1}(-1)^{-q_1}u^{2p_1}}f(\rho+4q_1,\mu)
,\ee
\be \label{modelKraising1} K^+f(\rho,\mu)=(\frac{\rho}{4}-\frac{\mu}{2k_1}-\frac{d}{2}+\frac12)_{q_1p_2}(-\frac{\rho}{4}-\frac{\mu}{2k_1}+\frac{d}{2}+\frac12)_{q_1p_2}\times\ee
\[(-1)^{q_1p_2}(-\frac{\mu}{k_1})_{2q_1p_2}2^{p_1q_2}(-\frac{\mu}{2k_2}+\frac{b}{2}-\frac{c}{2}+\frac12)_{p_1q_2}(-\frac{\mu}{2k_2}-\frac{b}{2}+\frac{c}{2}+\frac12)_{p_1q_2}
 f(\rho,\mu-2p_1q_2),\]
\be \label{modelKlowering1} K^-f(\rho,\mu)=\ee
\[\frac{(\frac{\rho}{4}+\frac{\mu}{2k_1}+\frac{d}{2}+\frac12)_{q_1p_2}(-1)^{q_1p_2}(-\frac{\rho}{4}+\frac{\mu}{2k_1}-\frac{d}{2}+\frac12)_{q_1p_2}2^{p_1q_2}(\frac{\mu}{2k_2}-\frac{b}{2}-\frac{c}{2}+\frac12)_{p_1q_2}(\frac{\mu}{2k_2}+\frac{b}{2}+\frac{c}{2}+\frac12)_{p_1q_2}}{(\frac{\mu}{k_1}+1)_{2q_1p_2}}\]
\[\times f(\rho,\mu+2p_1q_2). \]  The space of rational functions is invariant under the  algebra action, but polynomials are not. 

\subsection{The special case $k_1=k_2=1$} \label{specialcase}
In the case $k_1=k_2=1$  we are in Euclidean space and our system has additional symmetry. In terms of Cartesian coordinates
$x=r\sin\theta_1\cos\theta_2,\ y=r\sin\theta_1\sin\theta_2,\ z=r\cos\theta_1, $
the Hamiltonian is
\be\label{cartham1} { H}=\partial_x^2+\partial_y^2+\partial_z^2+\frac{\alpha}{r}+\frac{\beta}{x^2}+\frac{\gamma}{y^2}+\frac{\delta}{z^2}.\ee
Note that any permutation of the ordered pairs $(x,\beta),(y,\gamma),(z,\delta)$  leaves the Hamiltonian unchanged. This leads to 
additional structure in the symmetry algebra.
The basic symmetries are (\ref{L24}), (\ref{L34}).
The permutation symmetry of the Hamiltonian shows that ${ I}_{xz},{ I}_{yz}$ are also symmetry operators, and 
${ L}_2= { I}_{xy}+{ I}_{xz}+{ I}_{yz}-(\beta+\gamma+\delta)$. The constant of the motion ${ K}_0$ is 2nd order:
$${K}_0=-\frac18{ I}_{yz}-\frac{1}{16}{ L}_3+\frac{1}{16}({L}_2+\beta+\gamma+\delta)=-\frac{1}{32}({ I}_{yz}-{ I}_{xz}),$$
and ${ J}_0$ is 4th order:
\be\label{J0cart1}  J_0=-4 M_3^2-\frac12\{(\{x,\partial_x\}+\{y,\partial_y\}+\{z,\partial_z\})^2,\frac{\delta}{z^2}\}+2H\left(I_{xz}+I_{yz}-(\beta+\gamma+\frac34)\right)+5\frac{\delta}{z^2}+\frac{\alpha^2}{2},\ee
\be\label{laplace31} { M}_3=\frac12\{(y\partial_z-z\partial _y),\partial_y\}-\frac12\{(z\partial_x-x\partial_z),\partial_x\}-z\left(\frac{\alpha}{2r}+\frac{\beta}{x^2}+\frac{\gamma}{y^2}+\frac{\delta}{z^2}\right).\ee
(We have checked explicitly that  the symmetries given by these Cartesian expressions agree exactly with the special cases $k_1=k_2=1$ of the spherical coordinate expressions for these operators derived in the preceding sections.)
If $\beta=\gamma=\delta=0$ then ${ M}_3$ is itself a symmetry operator.

The symmetries ${ H}, { L}_2,{ L}_3,{ J}_0,{ K}_0$ form a generating (rational) basis for the symmetry operators.
Under the transposition $(x,\beta)\leftrightarrow(z,\delta)$ this basis is mapped to an alternate basis ${ H}, { L}'_2,{ L}'_3,{ J}'_0,{ K}'_0$:
$$ { L}'_2 = { L}_2,\quad  { L}'_3=-16{ K}_0+\frac12{ L}_2-\frac12{ L}_3+\frac{\beta+\gamma+\delta}{2},\quad { K}'_0=\frac12{ K}_0+\frac{1}{64}{ L}_2-\frac{3}{64}{ L}_3+\frac{1}{64}(\beta+\gamma+\delta),$$ 
\be\label{R'idents1} { R}_1'=[{ L}_2,{ J}_0'],\ { R}_2'= [{ L}_3',{K}_0']=-[L_3,K_0]=-{ R}_2,\
 { R}_3'=[{J}_0',{ K}_0']=-\frac{1}{32}{ R}_1'+\frac{1}{32}[{ L}_3,{ J}_0'].\ee
All of the identities derived earlier  hold for the primed symmetries. It is easy to see that the ${ K}'$ symmetries are simple polynomials in the ${ L},{ K}$ symmetries already constructed, e.g., ${ K}'_1=[{ L}'_3,{ K}'_0]=-{ K}_1$.
However, the ${J}'$ symmetries are new. In particular,
\be\label{newJ01} { J}'_0=-4 M_1^2-\frac12\{(\{x,\partial_x\}+\{y,\partial_y\}+\{z,\partial_z\})^2,\frac{\beta}{x^2}\}+2H\left(I_{xz}+I_{xy}-(\gamma+\delta+\frac34)\right)+5\frac{\beta}{x^2}+\frac{\alpha^2}{2},\ee
\be\label{laplace11} { M}_1=\frac12\{(y\partial_x-x\partial _y),\partial_y\}-\frac12\{(x\partial_z-z\partial_x),\partial_z\}-x\left(\frac{\alpha}{2r}+\frac{\beta}{x^2}+\frac{\gamma}{y^2}+\frac{\delta}{z^2}\right).\ee
Note that the transposition $(y,\gamma)\leftrightarrow(z,\delta)$ does not lead to anything new. Indeed, under the symmetry we would obtain a constant of the motion
$$ { J}_0''=-4 M_2^2-\frac12\{(\{x,\partial_x\}+\{y,\partial_y\}+\{z,\partial_z\})^2,\frac{\gamma}{y^2}\}+2H\left(I_{xy}+I_{yz}-(\beta+\delta+\frac34)\right)+5\frac{\gamma}{y^2}+\frac{\alpha^2}{2},$$
\be\label{laplace21} { M}_2=\frac12\{(yz\partial_y-y\partial _z),\partial_z\}-\frac12\{(y\partial_x-x\partial_y),\partial_x\}-y\left(\frac{\alpha}{2r}+\frac{\beta}{x^2}+\frac{\gamma}{y^2}+\frac{\delta}{z^2}\right),\ee
but it is straightforward to check that
\be\label{Jident1}{ J}_0+{ J}'_0+{J}''_0 =-\frac12H+\frac{\alpha^2}{2},\ee
so that the new constant depends linearly on the previous constants.
For further use, we remark that under the  symmetry $(x,\beta)\leftrightarrow(y,\gamma)$ the constants of the motion  ${ L}_2,{ L}_3,{ J}_0,{ J}_1$ 
are invariant, whereas ${ K}_0,{ K}_1$ change sign. The action on ${ J}_0'$ is more complicated: 
${ J}'_0\longrightarrow { J}_0''=\frac{\alpha^2}{2}-\frac12H-{ J}_0-{ J}_0'$.

Note here that, with respect to the standard inner product on Euclidean space, the $L$ symmetries and $J_0,J'_0,K_0$ are all formally self-adjoint whereas the commutators $K_1,J_1$ are skew-adjoint. 
$$ H^*=H,\ L_2^*=L_2,\ L_3^*=L_3,\  K_0^*=K_0,\ J_0^*=J_0,\ (J_0')^*=J_0'\ J_1^*=-J_1,\ K_1^*=-K_1.$$
Further the commutators $J_2+2J_1, K_2+K_1$ are formally self-adjoint. Hence $J_2^*=J_2+4J_1$, $K_2^*=K_2+2K_1$. Finally, when acting on a generalized eigenbasis
 the raising and lowering operators satisfy
\be\label{adjoint} (J^+)^*=J^-(1+\frac{4}{\rho}),\quad (J^-)^*=J^+(1-\frac{4}{\rho}),\quad (K^+)^*=K^-(1+\frac{2}{\mu}),\quad (K^-)^*=K^+(1-\frac{2}{\mu}).\ee
Note also the operator identities on a generalized eigenbasis:
$$K^+(\mu-2)=\mu K^+,\ K^-(\mu+2)=\mu K^-,\ J^+(\rho-4)=\rho J^+, \ J^-(\rho+4)=\rho J^-.$$
These adjoint properties and symmetry with respect to permutations can be used to greatly simplify the determination of structure equations. For example, one can see that it isn't possible for a linear term in the generator $J_0'$ to be invariant under the transposition $(x,\beta)\leftrightarrow(y,\gamma)$  and if such a term changes sign it must be proportional to $J_0+2J_0'+\frac12 H-\frac{\alpha^2}{2}$.

For 
$${ Q}=({ L}_3-{ L}_2-\delta)^2-2(2\delta+1){ L}_2-L_3+2\delta+1, $$
and  $k_1=k_2=1$
the relevant rational structure equations simplify to 
$$[J_1,K_1]Q=\{J_1,K_2\}(2L_2-2L_3+2\delta+1)+\{J_2,K_1\}(L_2-L_3-\delta-1),$$
$$[J_1,K_2]Q=-\{J_1,K_1\}L_3(2L_2-2L_3+2\delta+1)+\{J_2,K_2\}(L_2-L_3-\delta-1),$$
$$4K_0(L_3+1)=K_2-K_1-\frac14(L_2-\delta)(\gamma-\beta),\quad [K_1,L_3]=4K_2+4K_1,\ [L_2,J_0]=J_1,$$
$$2J_0(4L_2+3)=J_2-2J_1-4(L_3-\delta+\frac14)(H-\alpha^2),\quad [J_1,L_2]=2J_2+4J_1,\  [L_3,K_0]=K_1.$$

In the paper \cite{DASK2011}, Tanoudis and Daskaloyannis show that the quantum symmetry algebra generated by the 6 functionally dependent symmetries 
${ H}, { L}_2,{ L}_3,{ J}_0,{ K}_0$ and ${ J}'_0$  closes polynomially, in the sense that all double commutators of the generators are 
again expressible as noncommuting  polynomials in the generators. However more is true. All formally self-adjoint symmetry operators, such as $\{K_1,J_1\}$, $K_1^2$, $J_1^2$, $[K_1,J_1]$ are expressible as symmetrized polynomials in the generators. Further there is a polynomial relation among the 6 generators.
To see this let's first consider the symmetries $K_1^2$, $J_1^2$.  We expect these order 6 and order 10 symmetries to be expressed as polynomials in 
the generators. Using the corresponding expression in the classical case to get the highest order terms, as well as the adjoint and permutation symmetry 
constraints, we find the expansions to take the form
\be\label{K1ident} K_1^2+8\{L_3+5,K_0^2\}-4(\beta-\gamma)K_0(L_2-\delta)-C(L_2,L_3)=0,\ee 
$$C(L_2,L_3)=\frac14\gamma+\frac14\beta +\frac14\delta -\frac18L_3L_2\delta +\frac14L_3L_2\beta +\frac14L_3\delta \beta +\frac14L_2\delta \beta +\frac14L_3L_2\gamma$$
$$+\frac14L_3\delta \gamma+\frac14L_2\delta \gamma-\frac18L_3\beta \gamma+\frac14L_2\beta \gamma-\frac12L_3+\frac14L_2-\frac{7}{16}L_3^2+\frac{1}{16}L_3^3+
\frac18L_2^2 +\frac14\delta \beta$$
 $$ +\frac14\delta \gamma-\frac14\beta \gamma+\frac18\delta ^2+\frac18\beta ^2+\frac18\gamma^2-\frac18\delta ^2\beta -\frac18\delta \beta ^2-\frac18\delta ^2\gamma
-\frac18\delta \gamma^2 +\frac14\delta \beta \gamma+\frac18L_3\gamma$$
 $$+\frac18L_3L_2-\frac18L_3^2L_2+\frac{1}{16}L_3L_2^2+\frac18L_3\beta +\frac18L_3\delta +\frac14L_2\beta 
-\frac14L_2\delta+\frac14L_2\gamma-\frac18L_3^2\delta$$ $$ +\frac{1}{16}L_3\delta ^2-\frac18L_3^2\beta -\frac18L_2^2\beta +\frac{1}{16}L_3\beta ^2-\frac18L_2\beta ^2
-\frac18L_3^2\gamma-\frac18L_2^2\gamma+\frac{1}{16}L_3\gamma^2-\frac18L_2\gamma^2,$$ 
\be\label{J1ident}J_1^2+\{8L_2+38,J_0^2\}+4J_0\left(4HL_3+(1-4\delta)H-4\alpha^2L_3+\alpha^2(4\delta-1)\right)-D(H,L_2,L_3)=0, \ee
$$D(H,L_2,L_3)=5a^4-8L_3a^4+296E^2L_3+80H^2L_3^2+4a^4L_2-156H^2L_2-560H^2L_2^2-8a^4\delta +216H^2\delta +$$
$$80H^2\delta ^2-16HL_3a^2+32HL_3^2a^2+64H^2L_3^2L_2-72Ha^2L_2+288H^2L_3L_2+32Ha^2L_2^2-128H^2L_3L_2^2$$
$$-160H^2L_3\delta +224H^2L_2\delta -128H^2L_2^2\delta -48Ha^2\delta +64H^2L_2\delta ^2+32Ha^2\delta ^2+64H^2L_2^3-64HL_3a^2L_2-64HL_3a^2\delta $$
$$-64Ha^2L_2\delta -128H^2L_3L_2\delta -74Ha^2+229H^2.$$
Here $C$ is symmetry of order at most 6, and $D$ is a symmetry of order at most 8.

Using the fact that $J_0'$ must be formally self-adjoint and computing $[K_0, J_1]+4\{K_0,J_0\}$ on a generalized basis of eigenvectors 
$X[\mu,\rho]\equiv \Xi_{n,m,p}$ we define $J$  by
\be\label{buildingblocks}JX[\mu,\rho]= J_{pp}X[\mu-2,\rho-4]+ J_{pm}X[\mu-2,\rho+4]+ J_{mp}X[\mu+2,\rho-4]+ J_{mm}X[\mu+2,\rho+4]\ee
$$+ J_{zp}X[\mu,\rho-4]+ J_{zm}X[\mu,\rho+4]+ J_{pz}X[\mu-2,\rho]+ J_{mz}X[\mu+2,\rho]+ J_{zz}X[\mu,\rho],$$
$$ J_{pp}X[\mu-2,\rho-4]=\{K^+,J^+\} P_{++}(\mu,\rho)X[\mu,\rho],\  J_{pm}X[\mu-2,\rho+4]=\{K^+,J^-\} P_{+-}(\mu,\rho)X[\mu,\rho],$$
$$ J_{mp}X[\mu+2,\rho-4]=\{K^-,J^+\} P_{-+}(\mu,\rho)X[\mu,\rho],\ J_{mm}X[\mu+2,\rho+4]=\{K^-,J^-\} P_{--}(\mu,\rho)X[\mu,\rho],\ $$
$$ J_{zp}X[\mu,\rho-4]=J^+P_{0+}(\mu,\rho)X[\mu,\rho],J_{zm}X[\mu,\rho+4]=J^-P_{0-}(\mu,\rho)X[\mu,\rho],$$
$$ J_{pz}X[\mu-2,\rho]=K^+P_{+0}(\mu,\rho)X[\mu,\rho],J_{mz}X[\mu+2,\rho]=K^-P_{-0}(\mu,\rho)X[\mu,\rho], $$
$$J_{zz}X[\mu,\rho] =P_{00}(\mu,\rho]X[\mu,\rho].$$
In particular, 
$$ P_{--}(\mu,\rho)=\frac{-1}{2((\rho-2\mu)^2+4\delta+3)\rho(\rho+2)\mu(\mu+1)},\ P_{++}(\mu,\rho)=\frac{-1}{2((\rho-2\mu)^2+4\delta+3)\rho(\rho-2)\mu(\mu-1)},\ $$
$$ P_{+-}(\mu,\rho)=\frac{-1}{2((\rho+2\mu)^2+4\delta+3)\rho(\rho+2)\mu(\mu-1)},\ P_{-+}(\mu,\rho)=\frac{-1}{2((\rho+2\mu)^2+4\delta+3)\rho(\rho-2)\mu(\mu+1)}.\ $$
$$P_{0+}(\mu,\rho)=\frac{\beta-\gamma}{16\rho(\rho-2)(\mu+1)(\mu-1)},\ P_{0-}(\mu,\rho)=\frac{\beta-\gamma}{16\rho(\rho+2)(\mu+1)(\mu-1)},$$
$$P_{+0}(\mu,\rho)=-\frac{H-\alpha^2}{(\rho+2)(\rho-2)\mu(\mu-1)},\ P_{-0}(\mu,\rho)=-\frac{H-\alpha^2}{(\rho+2)(\rho-2)\mu(\mu+1)},$$
$$P_{00}(\mu,\rho)=\frac{(\beta-\gamma)(H-\alpha^2)(\rho^2+4\mu^2+\delta-\frac54)}{16(\mu^2-1)(\rho^2-4)}.$$
Then we can derive the polynomial identities
\be\label{princident1} [K_0,J_1]+4\{K_0,J_0\}-\frac{\gamma-\beta}{2}J_0+4(H-\alpha^2)K_0-\frac{\gamma-\beta}{4}(H-\alpha^2)+\{2(L_2-L_3+\delta+\frac12),J\}=0.\ee
\be\label{princident2} \{K_1,J_1\}+4\{-L_2-L_3+\delta+3,\{K_0,J_0\}\}+(\beta-\gamma)\{\frac32 L_2-\frac12 L_3+\frac{\delta}{2}+1,J_0\}\ee
$$-(H-\alpha^2)\{-4L_2+12L_3+4\delta+12,K_0\}+4\{Q,J\}+\frac12(\beta-\gamma)(H-\alpha^2)(4L_2+4L_3-20\delta+8)=0,$$
expressing the double commutator $[K_0,J_1]$ and the symmetrized product of commutators $\{K_1,J_1\}$ in terms of the generators.
We guess that $ J=\zeta(J_0+2J_0'+\frac12 H-\frac{\alpha^2}{2})$
for some scalar $\zeta$. A direct computation gives
$\zeta=-8$ and verifies the identities. Similar methods can be used to show polynomial closure of all terms of the form $\{[S_1,S_2],[S_3,S_4]\}$ where $S_1,\cdots, S_4$ are generators, but the details are complicated and we do not provide them here.

We conclude by sketching the determination of the algebraic relation between the 6 generators. The key here is the expression of all operators in terms of their action on a generalized basis of eigenvectors $X[\mu,\rho]$. Note that each of the operators $J^2,\{K_0^2,J_0^2\}, \{K_0,J_0\}^2, \{J,\{K_0,J_0\}\}$ can be expanded in the eigenbasis with only terms 
$$X[\mu\pm 4,\rho\pm 8],\ X[\mu\pm 4,\rho\mp 8],\ X[\mu\pm 4,\rho],\  X[\mu,\rho\pm 8], \ X[\mu\pm 4,\rho],\ X[\mu\pm 2,\rho],\ X[\mu,\rho\pm 4],\ X[\mu ,\rho].$$
Furthermore, the first differential operator is of order 8, the second and third are of order 12 and the fourth is of order 10. From these facts and the adjoint properties of the operators it is fairly straightforward to conclude that the  operator 
$$M_1\equiv  \{Q,J^2\}+32\{K_0,J_0\}^2-16\{L_2+L_3-\delta-1,\{J,\{K_0,J_0\}\}$$ is the unique  minimal order linear combination of these symmetries 
such that all terms in
$$X[\mu\pm 4,\rho\pm 8],\ X[\mu\pm 4,\rho\mp 8]$$
are zero. Next we compute the coefficients of $X[\mu,\rho\pm 8], \ X[\mu\pm 4,\rho]$ in the action of $M_1$ and show that these can canceled by an operator of the 
form
$ M_2\equiv \{ Q_1, K_0^2\}+\{Q_2,J_0^2\},$
where $Q_1$ is a polynomial in $H, L_2,L_3$ of order at most 2 and $Q_2$ is of order at most 4. Then $M_1+M_2$ has only terms in $ X[\mu\pm 2,\rho],\ X[\mu,\rho\pm 4],\ X[\mu ,\rho]$. Continuing in this way, we can cancel these remaining terms by an operator of the form 
$M_3\equiv \{Q_3,K_0\}+\{Q_4,J_0\}+Q_5$
where $Q_3$ is a polynomial in $H, L_2,L_3$ of order at most 5, $Q_4$ is of order at most 4, and $Q_5$ is of order at most 6. Thus we have the nontrivial differential operator  identity of order 12:
$$M_1+M_2+M_3\equiv 0$$
which is quadratic in $J_0'$. This is the algebraic relation that we seek. The final expression is too lengthy to list in a paper.

\section{Conclusions and Outlook}
One conclusion that we can reach from our analysis of the these extended Kepler-Coulomb systems is that for any rational pair $k_1,k_2$ the 3-parameter 
potential is never
just the restriction of the 4-parameter potential obtained by setting $\delta=0$. The 3-parameter system always has additional symmetries not inherited from 
the 4-parameter system. 
Note that  the $K$ raising and lowering operators fix $\rho$ and raise and lower $\mu$ by $\pm 2p_1p_2$ for each system, so the $K$ operators in the 3-parameter
 case are simply the restricted $K$ operators from the 4-parameter 
case. However, the 3-parameter $J$ operators fix $\mu$ and change $\rho$ by $\pm 2q_1$,  whereas the 4-parameter $J$ operators fix $\mu$ and change $\rho$ by 
$\pm 4q_1$, In essence, the 4-parameter $J$ operators become perfect squares upon restriction to $\delta=0$. This changes the structure of the symmetry algebra.

More generally, we have demonstrated that the  recurrence relation method developed in \cite{KKM10c} for proving superintegrability 
and determining the structure equations for
 families of  2D quantum superintegrable  systems can be extended to the 3D case. The construction appears to be quite general and not restricted to 
 Kepler-Coulomb analogs. For these higher order superintegrable systems it appears that algebraic closure is the norm. For polynomial closure, 
extra symmetry is needed. 
  We have not proved in all cases that there do not exist other generators of lower order but, if they exist, they must also be obtainable in 
terms of recurrence relations of hypergeometric functions.
A crucial role is played by the raising and lowering operators. They are not defined independent of eigenbasis and are not even symmetries, but
 all symmetries are built from them. The two variable models introduced here show promise in uncovering properties of rational orthogonal special functions
analogous to properties of orthogonal polynomials related to 2nd order superintegrable systems \cite{KMPost, KMP2011}

\section*{Acknowledgment} This work was partially supported by a grant from the Simons Foundation (\# 208754 to Willard Miller, Jr).

\end{document}